\global\def\draftcontrol{0}

   \def\versionno{NP quant geometry II}

\catcode`\@=11

\expandafter\ifx\csname draftcontrol\endcsname\relax\global\def\draftcontrol{0} 
\fi 

{\count255=\time\divide\count255 by 60 
\xdef\hourmin{\number\count255} 
\multiply\count255 by-60\advance\count255 by\time 
\xdef\hourmin{\hourmin:\ifnum\count255<10 0\fi\the\count255}} 
\def\draftdate{\number\month/\number\day/\number\year\ \ \ \hourmin } 

\newcommand\makepapertitle{\par

  \begingroup 
    \renewcommand\thefootnote{\@fnsymbol\c@footnote}%
    \def\@makefnmark{\rlap{\@textsuperscript{\normalfont\@thefnmark}}}%
    \long\def\@makefntext##1{\parindent 1em\noindent 
            \hb@xt@1.8em{%
                \hss\@textsuperscript{\normalfont\@thefnmark}}##1}%
     \newpage 
     \global\@topnum\z@   
     \@makepapertitle 
     \thispagestyle{empty}\@thanks 
  \endgroup 
  \setcounter{footnote}{0}%
  \global\let\thanks\relax 
  \global\let\makepapertitle\relax 
  \global\let\@makepapertitle\relax 
  \global\let\@thanks\@empty 
  \global\let\@author\@empty 
  \global\let\@date\@empty 
  \global\let\@title\@empty 
  \global\let\title\relax 
  \global\let\author\relax 
  \global\let\date\relax 
  \global\let\and\relax 
  \def\version{\let\version\@version\@gobble} 
} 
\def\@makepapertitle{%
  \newpage 
   \ifnum\draftcontrol=1 {} 
   \version\versionno 
   \vskip 5.5em%
   \else 
   \hfill\hbox to 3cm {\parbox{4.5cm}{\@pubnum}\hss}%
   \vskip 6.5em%
   \fi 
   \begin{center}%
   \let \footnote \thanks 
      {\hskip -0\textwidth \hbox to 1\textwidth%
        {\centerline{\Large\bf{\noindent\@title}}}}%
     \vskip 2em%
     {\normalsize
       \lineskip .5em%
       \begin{tabular}[t]{c}%
         \@author 
       \end{tabular}\par}%
     \vskip 1.5em%
     {\@bstract}%
     \end{center}%
     \vfill
     \@date%
     \vskip 1.5em%
   \par 
} 

\gdef\@pubnum{} 
\def\pubnum#1{%
  \gdef\@pubnum{#1}} 

\gdef\@bstract{} 
\def\Abstract#1{%
  \gdef\@bstract{%
   \parbox{\textwidth-0pc}{%
   \centerline{\bf Abstract}\penalty1000 
   \noindent
   \renewcommand\baselinestretch{1.0} 
   {#1}}} 
} 

\gdef\@email{}
\def\email#1{%
   \gdef\@email{%
   Email: {\tt #1}}
}

\def\ps@paper{\let\@mkboth\@gobbletwo%
     \ifnum\draftcontrol=1 
        \def\@oddfoot{\hbox to \textwidth{\tiny \versionno \hfil\tiny\draftdate}%
        \hskip -\textwidth \hbox to \textwidth{\hfil\rm\thepage\hfil}}%
     \else\def\@oddfoot{\hbox to \textwidth{\hfil\rm\thepage\hfil}} 
     \fi 
     \let\@evenfoot\@oddfoot 
} 

\def\body{\clearpage 
          \pagestyle{paper} 
        } 
\newenvironment{acknowledgments}{%
\vskip 3.25ex 
\addcontentsline{toc}{section}{Acknowledgments}
\noindent {\bf Acknowledgments} 
} 

\def\@version#1{\ifnum\draftcontrol=1 
\typeout{}\typeout{#1}\typeout{} 
\vskip3mm\centerline{\hbox{\fbox{\normalsize{\tt DRAFT -- #1 -- } 
                   {\draftdate}}}}\vskip3mm 
\fi} 
\let\version\@version 
\long\def\eqlabel#1{\ifnum\draftcontrol=1 
                    \tag@false  
                    \tag*{(\theequation) \hbox to -0.2cm{\hspace{0cm}\small{#1}\hss}} 
                    \refstepcounter{equation}  
                    \edef\@currentlabel{\theequation} 
                    \ltx@label{#1}          
                    \else 
                    \label{#1} 
                    \fi 
                    } 
\let\st@bibitem\@bibitem 
\let\st@lbibitem\@lbibitem 
\ifnum\draftcontrol=1 
  \def\@bibitem#1{%
    \st@bibitem{#1}\a@@label{#1}\ignorespaces} 
  \def\@lbibitem[#1]#2{%
    \st@lbibitem[#1]{#2}\a@@label{#2}\ignorespaces} 
  \def\a@@label#1{%
    \gdef\a@lab{\smash{\normalfont\small#1}} 
    \ifvmode 
      \if@inlabel 
        \global\setbox\@labels\hbox{%
          \llap{\a@lab\let\a@lab\relax 
                \kern\@totalleftmargin\kern\marginparsep}%
          \box\@labels}%
      \fi 
    \fi} 
\fi 

\documentclass[12pt,letterpaper]{article} 

\usepackage{amsmath,bm,amsfonts,amssymb,array,calc,amsthm,rotating,amscd}
\usepackage{epsfig,psfrag} 
\usepackage{rotating}
\usepackage{amscd}
\usepackage{graphicx}
\usepackage{color}
\usepackage[colorlinks=false]{hyperref}

\renewcommand\baselinestretch{1.25} 
\renewcommand\section{\@startsection {section}{1}{\z@}%
                                   {-3.5ex \@plus -1ex \@minus -.2ex}%
                                   {2.3ex \@plus.2ex}%
                                   {\normalfont\large\bfseries}} 
\renewcommand\subsection{\@startsection{subsection}{2}{\z@}%
                                   {-3.25ex\@plus -1ex \@minus -.2ex}%
                                   {1.5ex \@plus .2ex}%
                                   {\normalfont\normalsize\bfseries}} 
\renewcommand\subsubsection{\@startsection{subsubsection}{3}{\z@}%
                                   {-3.25ex\@plus -1ex \@minus -.2ex}%
                                   {1.5ex \@plus .2ex}%
                                   {\normalfont\normalsize\it}} 
\renewcommand\paragraph{\@startsection{paragraph}{4}{\z@}%
                                   {-1.75ex\@plus -1ex \@minus -.2ex}%
                                   {1ex \@plus .2ex}%
                                   {\normalfont\normalsize\bf}} 
\renewcommand\subparagraph{\@startsection{subparagraph}{5}{\z@}%
                                   {-1.25ex\@plus -0ex \@minus -.2ex}%
                                   {-2ex \@plus .2ex}%
                                   {\normalfont\normalsize\it}}

\numberwithin{equation}{section}

\long\def\@makecaption#1#2{%
  \vskip\abovecaptionskip
  \sbox\@tempboxa{{\bf #1:} #2}%
  \ifdim \wd\@tempboxa >\hsize
    {\small\bf #1:} {\small #2}\par
  \else
    \global \@minipagefalse
    \hb@xt@\hsize{\hfil\box\@tempboxa\hfil}%
  \fi
  \vskip\belowcaptionskip}

\renewcommand*\l@section[2]{%
  \ifnum \c@tocdepth >\z@
    \addpenalty\@secpenalty
    \addvspace{.5em \@plus\p@}%
    \setlength\@tempdima{1.5em}%
    \begingroup
      \parindent \z@ \rightskip \@pnumwidth
      \parfillskip -\@pnumwidth
      \leavevmode \bfseries
      \advance\leftskip\@tempdima
      \hskip -\leftskip
      #1\nobreak\hfil \nobreak\hb@xt@\@pnumwidth{\hss #2}\par
    \endgroup
  \fi}
\renewcommand*\l@subsection{\addvspace{.0em \@plus\p@}\@dottedtocline{2}{1.5em}{2.3em}}
\renewcommand*\l@subsubsection{\addvspace{-.2em \@plus\p@}\@dottedtocline{3}{3.8em}{3.2em}}

\def\quantph#1{\href{http://xxx.arxiv.org/abs/quant-ph/#1}{{arXiv:quant-ph/#1}}}
\def\hepth#1{\href{http://xxx.arxiv.org/abs/hep-th/#1}{{arXiv:hep-th/#1}}}

\def\math#1{\href{http://xxx.arxiv.org/abs/math/#1}{{arXiv:math/#1}}}

\def\arxiv#1#2{\href{http://xxx.arxiv.org/abs/#1}{{arXiv:#1 [#2]}}}

\definecolor{refcol}{rgb}{0.2,0.2,0.8}
\definecolor{eqcol}{rgb}{.6,0,0}
\definecolor{purple}{cmyk}{0,1,0,0}

\gdef\@citecolor{refcol}
\gdef\@linkcolor{eqcol}
\def\colorlinkspurple{\gdef\@urlcolor{purple}}
\def\colorlinksblue{\gdef\@urlcolor{blue}}
\def\colorlinksred{\gdef\@urlcolor{red}}

\def\ie{{\it i.e.}}

\def\cf{{\it cf.}}

\def\revise#1       {\raisebox{-0em}{\rule{3pt}{1em}}%
                     \marginpar{\raisebox{.5em}{\vrule width3pt\ 
                     \vrule width0pt height 0pt depth0.5em 
                     \hbox to 0cm{\hspace{0cm}{%
                     \parbox[t]{4em}{\raggedright\footnotesize{#1}}}\hss}}}}

\def\cale         {{\cal E}}

\def\ii           {{\it i}}

\def\Re           {{\rm Re\hskip0.1em}}

\def\sqr#1#2{{\vcenter{\vbox{\hrule height.#2pt   
 \hbox{\vrule width.#2pt height#1pt \kern#1pt 
 \vrule width.#2pt}\hrule height.#2pt}}}}

\newcommand{\Z}{\mathbb Z}

\newcommand{\Bcal}{\mathcal B}

\newcommand{\Fcal}{\mathcal F}
\newcommand{\Tcal}{\mathcal T}
\newcommand{\Ocal}{\mathcal O}

\newcommand{\Ccal}{\mathcal C}

\newcommand{\Ncal}{\mathcal N}

\newcommand{\ep}{\epsilon}
\renewcommand{\t}{\tilde}
\newcommand{\h}{\hat}
\newcommand{\corr}[1]{\left<#1\right>}

\newcommand{\beq}{\begin{equation}}
\newcommand{\eq}{\end{equation}}
\newcommand{\req}[1]{(\ref{#1})}

\catcode`\@=12 

\begin{document} 


\title{Non-Perturbative Quantum Geometry II}

\pubnum{
SNUTP14-011
}
\date{October 2014}

\author{
Daniel Krefl$^a$   \\[0.2cm]
\it  $^a$ Center for Theoretical Physics, SNU, Seoul, South Korea\\
}

\Abstract{
The Nekrasov-Shatashvili limit of $\beta$-ensembles with polynomial potential and $\Ncal=2$ supersymmetric gauge theories in the $\Omega$-background is intimately related to complex one-dimensional quantum mechanics. Multi-instanton corrections in quantum mechanics, inferable from exact quantization conditions, imply additional non-perturbative corrections to the Nekrasov-Shatashvili free energies. Besides filling some of the gaps in previous derivations, we present analytic expressions for such additional non-perturbative corrections in the case of $SU(2)$ gauge theory expanded at strong coupling. In contrast, at weak coupling these additional non-perturbative corrections appear to be negligible. 
}

\makepapertitle

\body

\version\versionno

\vskip 1em


\section{Introduction}
In this work we continue the study of the non-perturbative completion of the Nekrasov-Shatashvili (for short NS) limit of $\beta$-ensembles with polynomial potential (and so at large $N$ refined topological string theory) and four dimensional supersymmetric gauge theories (with eight supercharges) in the $\Omega$-background, along the lines of our previous work \cite{K13}. As the NS limit relates gauge/string theories to complex one-dimensional quantum mechanical problems \cite{NS09,MM09,ACDKV11}, the main idea pursued in \cite{K13} is to make use of this correspondence to deduce the non-perturbative completion of the NS free energies from quantum mechanics. One should note that there are as well other attempts in the literature to infer the non-perturbative completion of the NS free energy, like for example \cite{KM13}. The non-perturbative completion we are considering here constitutes a minimal self-consistent non-perturbative sector, necessary to reproduce quantum mechanical instanton effects.

Most of the results of \cite{K13} were obtained at hand of an instructive example, namely the $\beta$-ensemble with cubic potential, known to be equivalent at large $N$ to the refined topological string on a corresponding Dijkgraaf-Vafa geometry. It was found that combining the large $N$ limit with the NS limit is subtile, \ie, besides the usual large $N$ limit with $t:=g_s\beta N$ fixed, leading to the refined topological string, one may take alternatively $\Ncal:=\beta N$ as t'Hooft coupling constant. Keeping $\beta N$ fixed at large $N$ is well-defined, because we have $\beta\rightarrow 0$ in the NS limit. In \cite{K13}, the limit with $\beta N$ fixed has been referred to as quantum limit, as it leads for the cubic to the well-known quantum mechanics of a (critical) double-well potential with symmetry breaking term. In particular, the quantum mechanical energy is directly linked to the ensemble free energy in this limit. Though it was hinted at that a derivation of this fact could be given along the lines of \cite{ACDKV11}, it was not made explicit. In section \ref{CubicSec} we will fill this gap.

A remark is in order. Both choices for the t'Hooft coupling constant lead in the NS limit to quantum mechanical systems. The case with $g_s\beta N$ fixed has been discussed extensively in \cite{ACDKV11}, leading to a notion of quantum special geometry from which the NS limit of the refined topological string free energy can be recovered. What we learned in \cite{K13} is, that there exists as well a sort of intermediate quantum geometry, arising for $\Ncal:=\beta N$ fixed. In a subsequent large $\Ncal$ limit, with $t=g_s\Ncal$ fixed, one recovers the original quantum geometry of \cite{ACDKV11}. The meaning of the geometry arising in the quantum limit will become more clear in section \ref{CubicSec}.

The relation between the quantum mechanical energy and the $\beta$-ensemble free energy mentioned above is of particular interest, as it holds non-perturbatively (this will also become clear in section \ref{EnsembleToQMsec}). In quantum mechanics, the non-perturbative corrections to the quantum mechanical energy due to instanton tunneling are captured by so-called exact quantization conditions (see the review series \cite{ZJJ04a,ZJJ04b} and references therein), mathematically derivable from resurgence (\cf, \cite{V83,DP99,DDP97}). For the quantum system arising from the cubic (on the anti-diagonal
slice), the exact quantization condition takes a particular simple form, being equivalent to the so-called Nekrasov-Shatashvili quantization condition known from $\Ncal=2$ supersymmetric gauge theories in the $\Omega$-background. It was found in \cite{K13} that the exact quantization condition can be solved analytically, order by order in an instanton counting parameter, leading to non-perturbative corrections to the quantum geometry, and so to the NS free energy. Therefore the notion of non-perturbative quantum geometry was coined in \cite{K13}. The results constitute a stringent test on any trans-series expansion of the $\beta$-ensemble with cubic potential, as in the quantum limit the non-perturbative contributions calculated in \cite{K13} must be reproduced. For instance, it would be very interesting to check against the proposal of \cite{SESV13}.

However, the work \cite{K13} has been incomplete in several ways. Besides some minor gaps in derivations mentioned already above, its main weakness is that the $\beta$-ensemble with cubic potential is remarkably special in the sense that there exists a very simple relation between the free energy of the ensemble and the energy of the corresponding quantum mechanical system. It is however hard to find other $\beta$-ensembles with a similarly simple relation, which one could investigate in a similar fashion. Hence, the question of the general validity and usability of the obtained results arise. 

Here, we show that the cubic discussed in \cite{K13} is in fact a very illustrative prototypical example for a, at first sight, very different class of models. Namely, four dimensional supersymmetric gauge theories with eight supercharges in the $\Omega$-background.  As an illuminating example for more complicated gauge theories we will investigate the case of pure $SU(2)$ gauge theory (without matter representations), and will indeed see that the non-perturbative structure of the NS limit of the gauge theory can be investigated along the lines of \cite{K13}. As the moduli space of $SU(2)$ gauge theory is more accessible than for instance the moduli space of the $\beta$-ensemble with cubic potential, the $SU(2)$ example allows us to investigate the behavior of the non-perturbative sector at different points in moduli space with little additional effort. In particular, we will learn that the presence of additional non-perturbative corrections depends on the point of expansion in Coloumb moduli space. Whereas the expansion at weak coupling is essentially exact, we have at strong coupling an additional non-perturbative sector along the lines of \cite{K13}. 

The outline is as follows. In the next section we will revisit the $\beta$-ensemble with polynomial potential, thereby filling some of the gaps left open in \cite{K13} and giving the conceptual foundation for the following sections. In section \ref{SU2sec}, we will start to discuss pure $\Ncal=2$ SU(2) gauge theory in the NS limit. Under quantization of the Coulomb parameter we recover the quantum mechanical Sine-Gordon model, which will be discussed in section \ref{SGmodel}. The non-perturbative completion of SU(2) gauge theory in the NS limit will be given in section \ref{NPsu2sec}. The appendix collects some more technical material needed in the main text. Namely, appendix \ref{SEQDerivation} gives additional details for the derivations in section \ref{EnsembleToQMsec}, appendix \ref{charNumbersExp} summarizes the calculation of the perturbative data for the Sine-Gordon model, respectively, $SU(2)$ via Mathieu's equation and in appendix \ref{appMasslessMultiplets} the contribution of massless vector/hyper-multiplets to the gauge theory free energy is evaluated in the NS limit.

As sort of a disclaimer, the reader should note that we give only limited warranty for phases and factors of two, though the conventions we picked appear to lead to an overall consistent scheme.

\section{The cubic revisited}
\label{CubicSec}
\subsection{From ensembles to quantum mechanics}
\label{EnsembleToQMsec}
Recall the partition function of an (holomorphic) $\beta$-ensemble of $N$ eigenvalues with potential $W(\lambda)$, 
\beq
\eqlabel{ZensembleDef}
Z_\Ccal(N,\beta,g_s):=\int_\Ccal [d\lambda] \Delta(\lambda)^{2\beta}\,e^{-\frac{\beta}{g_s} \sum_{i=1}^N W(\lambda_i)}\,,
\eq
with $\Delta(\lambda)$ the Vandermonde determinant $\Delta(\Lambda)=\prod_{i<j}^N(\lambda_i-\lambda_j)$, $\Ccal$ an integration path in the complex plane and $\beta$ some positive integer. We take $W(\lambda)$ to be a polynomial of degree $d$. There are $c=d-1$ critical points $z_*$ with $W'(z)|_{z_*}=0$. We assume that the critical points are non-degenerate and that $W''(z)|_{z_*}\neq 0$. The convergence properties of the partition function \req{ZensembleDef} depend on the choice of path $\Ccal$. There are $d$ angular sectors of convergence in the complex plane. Holomorphicity dictates that an integration path leading to a non-vanishing partition function must connect two different angular sectors of convergence. Hence, there can be at most $d(d-1)/2$ independent paths. Via taking transformation properties of \req{ZensembleDef} under eigenvalue rescalings into account, the number of independent paths can be further reduced, \cf, \cite{L03}. We can decompose a consistent integration path $\Ccal$ into such a basis, and so the partition function into topological sectors (or phases) there each eigenvalue is integrated along one of the basis paths. 

The insertion of an operator $\Ocal$ into the eigenvalue ensemble defines a correlator
\beq
\eqlabel{CorrDef}
\corr{\Ocal}:=\int_\Ccal[d\lambda] \Delta(\lambda)^{2\beta}\,\Ocal\, e^{-\frac{\beta}{g_s} \sum_{i=1}^N W(\lambda_i)}\,.
\eq
Trivially, one has for the empty insertion $\corr{.}:=\corr{1}=Z_\Ccal(N,\beta, g_s)$. Of particular interest for us is the so-called brane operator $\psi_h(x):=\prod_{i=1}^N(\lambda_i-x)^h$ with $h$ an positive integer, and its correlator $\corr{\psi_h(x)}$. We define
\beq
\eqlabel{Psidef}
\Psi_{k,h}(x):=e^{-\frac{k}{g_s}W(x)}\frac{\corr{\psi_h(x)}}{\corr{.}}\,.
\eq
It is convenient to define a correlator in the brane background via
$$
\corr{\Ocal}_h:=\corr{\psi_h(x)\Ocal}.
$$
In particular, we have $\corr{.}_h:=\corr{1}_h=\corr{\psi_h(x)}$. 

It can be shown that the operator \req{Psidef} satisfies a second order differential equation \cite{ACDKV11}. In particular, taking
\beq\eqlabel{Psi12paras}
\boxed{
\begin{split}
\Psi_1:\,\,\,\, k=\beta/2\,&,h=\beta\\
\Psi_2:\,\,\,\,k=1/2\,&, h=1
\end{split}
}\,,
\eq
and introducing a parameter $c_i$ with $c_1=\beta$ and $c_2=1$, we have (see appendix \ref{SEQDerivation}) 
\beq\eqlabel{UnifiedSE}
\Psi_i''(x)=\frac{c_i^2}{4g_s^2}\left(\left(W'(x)\right)^2- \frac{2g_s}{c_i} W''(x) \right)\Psi_{i}(x)-\frac{c_i^2e^{-\frac{c_i}{2g_s}W(x)}}{g_s\corr{.}} \sum_{i=1}^N\corr{\frac{W'(x)-W'(\lambda_i)}{x-\lambda_i}}_{c_i}\,.
\eq
As a side remark, the informed reader might notice that integrating over the insertion \req{Psidef} with parameters \req{Psi12paras} corresponds to bringing in a $1/2$-, respectively, $1/(2\beta)$-frational eigenvalue, and hence can be interpreted as a fractional 1-instanton amplitude, following \cite{D92,HHIKKNT04,MSW07}. What we will see below is that in fact the differential equation for the 1-instanton amplitude \req{UnifiedSE} implicitly determines all multi-instanton corrections to the free energy, at least in certain cases.

Since the potential $W$ is assumed to be a polynomial, \ie, $W(\lambda)=\sum_{i=0}^\infty t_i \lambda^i$, we can define a linear differential operator 
\beq\eqlabel{hfxDef}
\hat d(x):=-\frac{g_s}{\beta}\hat D(x) +c(x)\,,
\eq 
with $\hat D(x)$ a linear combination of first order derivates in the coupling constants $t_i$ (sometimes referred to as non-renormalizable moduli) and $c(x)$ a polynomial, yielding
$$
\hat f(x)\corr{.}_{c_i} = \sum_{i=1}^N\corr{\frac{W'(x)-W'(\lambda_i)}{x-\lambda_i}}_{c_i}\,.
$$
It is clear that $c(x)=N\,\frac{W'(x)-W'(0)}{x}$. Furthermore, the operator $\hat d$, viewed as a polynomial in $x$, is of degree $d-2$.

Under commuting $\hat d$ to the right, we can rewrite \req{UnifiedSE} as a multi-time dependent Schr\"odinger equation, where the $t_i$ are interpreted as time coordinates, as shown in appendix \ref{SEQDerivation}. The insertion $\Psi_1$ is special, as it allows a decoupling of the time-dependence via taking the limit $\beta\rightarrow 0$, yielding the time-independent Schr\"odinger equation (we also redefined $\hbar:=g_s$)
\beq\eqlabel{tiSEQgeneral}
\boxed{
\hbar^2\,\Psi_1''(x)=\,\left(\left(W'(x)\right)^2- \mathfrak f(x) \right)\Psi_{1}(x)
}\,,
\eq
with $\mathfrak f(x):=\hbar\left(W''(x)+2\mathfrak c(x)+\mathfrak d(x)\right)$,
\beq\eqlabel{fQMDef}
\begin{split}
\mathfrak c(x):=&\lim_{\beta\rightarrow 0}\,\beta N\, \frac{W'(x)-W'(0)}{x}\,,\\
\mathfrak d(x):=&\,\hbar\lim_{\beta\rightarrow 0}\beta\, \h D(x) \,\Fcal_\Ccal(N;\beta, \beta\,\hbar/2)\,.\\
\end{split}
\eq
and $\Fcal_\Ccal:=\log\corr{.}$. Clearly $\mathfrak f(x)$ is a polynomial in $x$ of degree $d-2$. We can parameterize the coefficients as $\mu_n$. The energy $\cale$ of the quantum mechanical system can be identified to be given by 
\beq\eqlabel{EQMDef}
\cale=\mu_0=\hbar\,(W''(0)+2\mathfrak c(0)+\mathfrak f(0))\,.
\eq
Some remarks are in order. Firstly, the derivation of \req{tiSEQgeneral} does not involve a perturbative expansion or a particular choice of integration contour $\Ccal$. Hence, the relation between the $\beta$-ensemble free energy and the quantum mechanical energy $\cale$ given through \req{EQMDef} and \req{fQMDef} holds non-perturbatively. One should however keep in mind that the choice of integration path translates to a choice of boundary conditions for $\Psi_1$. Different boundary conditions will lead to non-perturbatively different energies $\cale$.  Secondly, in order to have a non-trivial ($N$ dependent) energy $\cale$, the limit $\lim_{\beta\rightarrow 0}\beta$ in \req{fQMDef} should be non-vanishing. The gaussian potential $W(x)=\frac{1}{2}x^2$ implies that this requires a large $N$ limit. In detail, we have $\left(W'(x)\right)^2=x^2$ and $W''(x)=1$. Hence,
$$
\sum_{i=1}^N \corr{\frac{W'(x)-W'(\lambda_i)}{x-\lambda_i}}_h=N\,,
$$
such that $\h D(x)=0$ and $c(x) = N$. The Schr\"odinger equation \req{tiSEQgeneral} becomes
\beq\eqlabel{HarmOscSE}
-\frac{\hbar^2}{2} \Psi_1''(x)+\frac{x^2}{2}\,\Psi_{1}(x)= \cale\, \Psi_{1}(x)\,,
\eq
with
\beq\eqlabel{HarmOsciEnergy}
\cale= \hbar\,\left(\frac{1}{2}+\lim_{\beta\rightarrow 0} \beta N\right)\,.
\eq
Clearly, a non-trivial energy requires that we take a large $N$ limit keeping either $\Ncal:=\beta N$ or $t:=\hbar\beta N$ fixed. The former limit has been first introduced in \cite{K13} and dubbed quantum limit, though a better name would be critical quantum limit, as will become more clear later. This limit yields for the gaussian potential the usual quantum mechanical harmonic oscillator with energy $\cale=\hbar\left(\Ncal+\frac{1}{2}\right)$, as is clear from \req{HarmOscSE} and \req{HarmOsciEnergy}. The latter limit is the usual large $N$ limit (with $g_s\ll 1$) leading to refined topological string theory (more precisely the Nekrasov-Shatashvili limit thereof, in our context). The Schr\"odinger equation \req{HarmOscSE} in the usual large $N$ limit can be interpreted as the quantum geometry of the deformed conifold, following \cite{ADKMV03,ACDKV11}. 

One should note that the NS limit of the usual large $N$ limit can also be reached in a two step process. First, one takes the quantum limit, and subsequently large $\Ncal$ (with $\hbar\ll 1$) holding 
\beq\eqlabel{tQuant}
\boxed{
t=\hbar\,\Ncal
}\,,
\eq 
fixed. The inverse is also true. Substituting \req{tQuant} after taking the NS limit, will result in the small $\hbar$ expansion of the quantum limit, at least perturbatively.

\paragraph{Example: The cubic}
For the eigenvalue ensemble with cubic potential discussed extensively in \cite{K13}, 
\beq
\eqlabel{cubicPotential}
W(x)=\frac{1}{3}x^3-\frac{\delta}{4}x\,,
\eq
we have
$$
\left(W'(x)\right)^2=\left(x^2-\frac{\delta}{4}\right)^2\,,\,\,\,\, W''(x)=2x\,.
$$
Furthermore,
$$
\sum_{i=1}^N \corr{\frac{W'(x)-W'(\lambda_i)}{x-\lambda_i}}_h=\sum_{i=1}^N\corr{x+\lambda_i}_h=\left(N\,x+\frac{4g_s}{\beta}\,\partial_\delta\right)\corr{.}_h\,.
$$
Hence,
$$
\h D(x)= -4\,\partial_\delta\,,\,\,\,\,\,c(x)=N\, x\,.
$$
Let us first consider the large $N$ limit with $\Ncal:=\beta N$ fixed. In this limit we have the time-independent Schr\"odinger equation
\beq\eqlabel{anHoscQMSE}
-\frac{\hbar^2}{2}\,\Psi_1''(x)+\left(\frac{1}{2}\left(x^2-\frac{\delta}{4}\right)^2 - \hbar\left( \Ncal+1\right)x \right)\Psi_{1}(x)= \frac{\hbar\,\cale_r}{2}\,\Psi_{1}(x)\,.
\eq
For 
\beq\eqlabel{CubicNcondition}
\Ncal=j-1\,,
\eq 
we recognize the quantum double-well with symmetry breaking term. In particular at $j=0$ we recover the (critical) double-well. The rescaled energy $\cale_r$ reads
$$
\boxed{
\cale_r:=\cale/\hbar=-4\hbar\frac{\partial\Fcal_Q(\Ncal,\hbar)}{\partial\delta}
}\,,
$$
with 
\beq\eqlabel{QlimitEnsembleDef}
\Fcal_Q(\Ncal,\hbar):=\lim_{\beta\rightarrow 0}\beta \Fcal_\Ccal(\Ncal/\beta,\beta,\beta\hbar/2)\,,
\eq
as conjectured in \cite{K13}. Taking instead the large $N$ limit with $t =\hbar\,\Ncal=\hbar\beta N$ fixed, we recover the Schr\"odinger equation of \cite{ACDKV11}, \ie,
$$
-\hbar^2\,\Psi_1''(x)+\left(\left(W'(x)\right)^2 +\mathfrak f(x) \right)\Psi_{1}(x)=0\,,
$$
with $\mathfrak f(x)=-2(t+\hbar)x+\mu_0(t)$ and $\mu_0(t)=-4\hbar^2\partial_\delta \Fcal_Q(t/\hbar,\hbar)$. In particular, for $t=-\hbar$, we recognize the quantum mechanics of a (non-critical) double-well potential.

\subsection{Semi-classical expansion}
\label{EnsembleClassicAsymptoticsSec}
In the gaussian case the partition function \req{ZensembleDef} can be evaluated exactly via Mehta's integral formula to be given by
\beq\eqlabel{Zgaussian}
Z_g(N,\beta, g_s)=(2\pi)^{N/2}\left(\frac{g_s}{\beta}\right)^{\frac{1}{2}\left(N+\beta N(N-1)\right)}\prod_{n=1}^N\frac{\Gamma(1+n\beta)}{\Gamma(1+\beta)}\,.
\eq
(The origin of the $g_s/\beta$ factor lies in a rescaling of the eigenvalues to match the gaussian $\beta$-ensemble with Mehta's integral.) For higher order potentials the evaluation of \req{ZensembleDef} is less straight-forward. However, in the $g_s\ll1$ limit (with $g_s\ll \beta$) an asymptotic expansion thereof can be found with relative ease. The reason being that holomorphicity allows us to deform the integration path $\Ccal$ to pass through the critical points $z_*$ in way such that near a critical point $z_*$ we can expand $W(z)$ as,
$$
W(z)=W(z_*)+w_2(z-z_*)^2+\Ocal\left((z-z_*)^3\right)\,,
$$
with $w_2=\frac{1}{2}W''(z)|_{z_*}\neq 0$, due to our original assumption on $W(z)$, and $w_2(z-z_*)^2$ real and positive. As for $g_s\ll 1$ the integrations in \req{ZensembleDef} localize infinitesimal close to the critical points, we can introduce local coordinates $y= c\sqrt{\frac{g_s}{\beta}}(z-z_*)$ with $c$ some constant such that at lowest order in $g_s$ the partition function \req{ZensembleDef} factorizes into gaussians with $N_i$ eigenvalues, where the distribution of eigenvalues with $\sum_{i=1}^c N_i=N$ is a priori arbitrary. Under factoring out the pure gaussian contributions, the higher order terms in $g_s$ are given by sums of normalized gaussian correlators, which can be evaluated explicitly, following \cite{KMT02,MS10,KW12} (see also \cite{K13} for a brief summary). Hence, we have that asymptotically
\beq\eqlabel{ZensembleAsymptotics}
Z_\Ccal(N,\beta,g_s)\xrightarrow{g_s\ll 1} Z(N;\beta,g_s)= e^{-\frac{\beta}{g_s}\sum_{i} N_i W(z_*^{(i)})}\left(\prod_{i=1}^c Z_g(N_i,\beta,g_s)\right) \left({\rm const.}+\Ocal\left(g_s\right)\right)\,.
\eq
We also define a perturbative free energy $\Fcal:=\log Z$. Clearly, $\Fcal$ does not depend on $\Ccal$ due to holomorphicity, \ie, $\Fcal_\Ccal$ with different choices of $\Ccal$ can only be distinguished non-perturbatively. 

The limit $\lim_{\beta\rightarrow 0}\beta\Fcal$ can be easily calculated from \req{ZensembleAsymptotics} making use of \req{Zgaussian}. For $\Ncal_i:=\beta N_i$ fixed, it is convenient to define a perturbative quantum free energy (or prepotential) as in \req{QlimitEnsembleDef},
\beq\eqlabel{FQensembleDef}
\Fcal_Q(\Ncal,\hbar):= \lim_{\beta\rightarrow 0}\beta\,\Fcal(\Ncal/\beta, \beta, \beta\hbar/2 ) \,,
\eq
and its derivative
$$
\Pi_{Q,i}(\Ncal):=\frac{1}{\hbar}\frac{\partial \Fcal_Q}{\partial\Ncal_i}\,.
$$
For the gaussian we simply have (\cf, \cite{KS13,K13})
\beq\eqlabel{GaussianPIB}
\hbar\,\Pi_Q(\Ncal)=\log\Gamma(1+\Ncal)+ \left(\Ncal+\frac{1}{2}\right)\log\left(\frac{\hbar}{2}\right)\,.
\eq
Combining \req{GaussianPIB} with \req{ZensembleAsymptotics} we infer the general expansion
\beq\eqlabel{PiBgeneral}
\hbar\,\Pi_{Q,i}(\Ncal)=-\frac{2 }{\hbar}\,W(z_*^{(i)})+\log\Gamma\left(1+\Ncal_i\right)-\left(\Ncal_i+\frac{1}{2}\right)\log\left(\frac{2}{\hbar}\right)+\Ocal\left(\hbar^0\right)\,.
\eq 
As the second and third term in \req{PiBgeneral} do not depend on the non-renormalizable moduli, we learn that $\h D(x) \Fcal_Q=\Ocal\left(1/\hbar\right)$ as $\hbar\rightarrow 0$. This will be of relevance for section \ref{pertquantgeo}. For $t_i:=\hbar \beta N_i$ fixed, we similarly define
\beq\eqlabel{FNSensembleDef}
\Fcal_{NS}(t,\hbar):=\lim_{\beta\rightarrow 0} \beta\, \Fcal(t/(\hbar\beta),\beta,\beta\hbar/2)\,,
\eq
and
$$
\Pi_{NS, i}(t):=\frac{\partial \Fcal_{NS}}{\partial t_i}\,.
$$
The reason why we refer to \req{FNSensembleDef} as NS free energy will become more clear in section \ref{SU2sec}. The $\Gamma$-function in \req{PiBgeneral} has to be understood as asymptotically expanded under substituting $\Ncal=t/\hbar$, using \req{LogGasymp}. Note that it is well-known that $\hat D(x)\Fcal_{NS}=\Ocal(1/\hbar^2)$. Clearly, as a classical asymptotic expansion $\Pi_{NS,i}(\hbar\, \Ncal)= \Pi_{Q,i}(\Ncal)$ and so $\Fcal_{NS}(\hbar\,\Ncal)=\Fcal_Q(\Ncal)$. 

As a side remark, one should note that for the cubic example sketched in the previous section one has to go onto the anti-diagonal slice in moduli space ($\Ncal_1=-\Ncal_2$) in order to enforce \req{CubicNcondition} in the semi-classical limit (for details see \cite{K13}).

\subsection{Perturbative quantum geometry}
\label{pertquantgeo}
In the semi-classical limit $\hbar\ll 1$ we can perform a WKB like Ansatz for the wave-function $\Psi_1$. In particular, at leading order in $\hbar$ we can identify the action integrals as (open) periods on a hyperelliptic curve
$$
\Sigma:\,y^2=\left( W'(x)\right)^2-\mathfrak f(x)\,,
$$
with canonical 1-form $dx\, y(x)$. One should note that $\Sigma$ is well-defined in both large $N$ limits (with $\hbar \ll 1$). In the usual large $N$ limit we have that $\mathfrak f = \Ocal\left(\hbar^0\right)$ and so the coefficients $\mu_n(t)$. Hence, the double zeros of $(W'(x))^2$ form brunch cuts and $\Sigma$ is classically smooth, forming the well-known large $N$ geometry of Dijkgraaf-Vafa. However, in the quantum limit of \cite{K13} we instead have that $\mathfrak f = \Ocal\left(\hbar\right)$ and so $\mu_n(\Ncal)$. The curve is classically singular. In particular, we must have that under the quantization \req{tQuant} the complex structure parameters of the curve $\Sigma$ rescale as 
\beq\eqlabel{muQuant}
\mu(t)\xrightarrow{t_i\rightarrow \hbar\, \Ncal_i} \hbar\, \mu(\Ncal)\,,
\eq
\ie, are quantized as well.

Following \cite{ACDKV11}, we can define a semi-classical quantum differential $dS$ on $\Sigma$ via taking all orders of the WKB expansion of $\Psi_1$, denoted as $\Psi_1^{\rm WKB}$, into account, \ie,
\beq\eqlabel{DefdS}
dS:=-\ii \hbar\, dx \,\partial_x \log \Psi^{\rm WKB}_1(x)\,.
\eq
(Where we made an arbitrary choice of momentum $\Psi_1^{WKB}(x)\sim e^{\frac{\ii}{\hbar}\int^x dS}$.)
The pair $(\Sigma,dS)$ is referred to as (perturbative or semi-classical) quantum geometry. Integrating $dS$ along a 1-cycle $\gamma$ of $\Sigma$  yields a so-called (perturbative) quantum period $\Pi_\gamma$. The quantum period is simply given by the phase $e^{\ii \phi_\gamma}$ picked up by $\Psi_1^{WKB}$ under analytic continuation along $\gamma$, \ie, 
$$
\Pi_\gamma(\mu):=\oint_\gamma dS= \hbar\, \phi_\gamma \,.
$$
This brings us to one of the main observations of \cite{ACDKV11} (see also \cite{MMM10,BMT11}). It is known from \cite{DV02a,DV02b} that in the usual large $N$ limit the classical A-periods around the cuts of $\Sigma_{NS}$ yield the corresponding filling fraction of ensemble eigenvalues. It appears that this generalizes to all orders in $\hbar$, \ie, we have 
\beq\eqlabel{EnsembleMirrorMaps}
\Pi_{\mathcal A_i,NS}(\mu) =\oint_{\mathcal A_i} dS =  t_i\,.
\eq 
Note that one should view \req{EnsembleMirrorMaps} as mirror maps, allowing to solve for the complex structure parameters $\mu_n(t)$ as a series in $\hbar$ and flat-coordinates $t_i$.

Furthermore, it has been observed that the integrability relation extends as well to higher orders in $\hbar$, \ie, there exists a quantum prepotential $\Fcal_{NS}$ with
$$
\Pi_{\mathcal B_i, NS}(t) =\oint_{\Bcal_{i}} dS= \frac{\partial\Fcal_{NS}}{\partial t_i}\,,
$$
where $\Fcal_{NS}$ corresponds to the ensemble free energy defined in \req{FNSensembleDef}. (In general, these observations remain to our knowledge unproofen so far. See however \cite{EM12} for the case of $\Ncal=2$ SU(2) gauge theory with four flavors and \cite{KPT14} for $\Ncal=2^*$ SU(2) gauge theory.) Substituting \req{tQuant} into the above periods leads to $\Pi_{\mathcal A_i,Q}$ and $\Pi_{\mathcal B_i,Q}$, where $\oint_{\mathcal A_i}$ should be understood as integration along the vanishing $\mathcal A_i$-cycle. Clearly, we have that $\phi_{\mathcal A_i}=\Ncal_i$, 
implying uniqueness of $\Psi_1^{WKB}$ under monodromies along the vanishing cycles for $\Ncal_i\in \Z$. In particular, $\Pi_{\mathcal A_i,Q}(\mu)=\hbar\, \Ncal_i$ should be seen as exact Bohr-Sommerfeld quantization conditions.

We can now give a perturbative interpretation of the quantum limit \req{FQensembleDef} introduced in \cite{K13} from a B-model topological string point of view. Namely, it simply corresponds to a quantization of the B-model target space geometry via the NS limit ($\beta\rightarrow 0$), combined with a quantization of the mirror A-model K\"ahler moduli via \req{tQuant} and so of the B-model complex structure moduli via \req{muQuant}. Classically, the quantization of moduli brings us into a region of moduli space where the target space becomes singular.

\subsection{Beyond semi-classics}
\label{EnsembleNPcompletion}
One should note that the expansions of the previous subsection are only classical asymptotic expansions and that it is in general not easy to improve on that with $\beta$-ensemble, respectively, matrix model techniques. The limit $\beta\rightarrow 0$ is however special, as due to the correspondence to quantum mechanics via relations \req{tiSEQgeneral}, \req{fQMDef} and \req{EQMDef}, the non-perturbative completion is, at least conceptually, clear.

Let us be more explicit what the non-perturbative completion is about. Recall from subsection \ref{EnsembleClassicAsymptoticsSec}, see \req{fQMDef}, that the complex structure parameters of the semi-classical quantum geometry $(\Sigma,dS)$ are given in terms of derivatives of the $\beta$-ensemble free energy in the non-renormalizable moduli. By derivation, this relation holds non-perturbatively, \ie, the $\mu_n$ are given in terms of $\Fcal_\Ccal$. Hence, beyond classical asymptotics, the $\mu_n$ will not be anymore series in $\hbar$, determined by $\Fcal$, but rather trans-series with some instanton counting parameter $\xi$, determined by $\Fcal_\Ccal=\Fcal+\Ocal(\xi)$, \ie, 
\beq\eqlabel{NPcomplexStructure}
\mu=\mu_p+\mu_{np}\,,
\eq
with $\mu_{np}$ corresponding to the non-perturbative corrections, as a series in $\xi$ (with coefficients series in $\hbar$, but not necessarily regular, \ie, terms singular in $\hbar$, like powers of $1/\hbar$ or $\log\hbar$, might occur). In turn, knowledge of $\mu_{np}$ allows us, via \req{fQMDef}, to infer $\Fcal_\Ccal$ up to some integration constant (more precisely up to a function independent of the non-renormalizable moduli).

As the perturbative quantum differential $dS$ defined in \req{DefdS} is a function of the complex structure parameters, the differential will be as well a trans-series beyond classical-asymptotics, \ie, we have a non-perturbative quantum geometry $(\Sigma, dS(\mu))$, and so non-perturbative quantum periods 
$$
\Pi_\gamma(\mu)=\Pi_\gamma(\mu_p+\mu_{np})=\hbar\, \phi_\gamma\,.
$$
It is important to keep the following in mind. Consider for instance $\Pi_{\mathcal A,NS}(\mu)=t$. Then, $t$ should be viewed as the non-perturbatively flat coordinate arising at large $N$, \ie, 
$$
t(\mu)=t(\mu_p)+\Ocal(\xi)=\hbar \beta N\,.
$$
Similarly, we have in the quantum limit $\Ncal(\mu)=\Ncal(\mu_p)+\Ocal(\xi)=\beta N$.

It remains to determine $\mu_{np}$. If we tune the moduli such that all $\mu_{n>0}$ are fixed, this is particulary easy, as $\mu_0=\cale$, and so the determination of $\mu_{np}$ translates to the calculation of the exact quantum mechanical energy $\cale$ (as was the case for the cubic on the anti-diagonal slice discussed in \cite{K13}). The split \req{NPcomplexStructure} reads for the energy $\cale=\cale_p+\cale_{np}$, with $\cale_{np}$ being the contribution of non-perturbative corrections due to instanton tunneling. For the calculation of the exact $\cale$ we can harvest well established quantum mechanical results, see \cite{DDP97,ZJJ04a,ZJJ04b} and references therein. In essence, leading to so-called exact quantization conditions (the  particular form of which depends on the model, point in moduli space, and choice of boundary conditions), the energy $\cale$ has to satisfy and which can be solved analytically for $\cale_{np}$ order by order in $\xi$, as illustrated at hand of the quantum limit of the cubic \req{cubicPotential} in \cite{K13}.

The general case, with all $\mu_n$ free, is less clear. However, we expect that resurgence techniques, along the lines of \cite{DP99,DDP97} may still be used to determine the non-perturbative corrections $\mu_{np}$. We leave this to follow-up works. Instead, here we will discuss a quite different model, which is however non-perturbatively solvable as the $\beta$-ensemble with cubic potential, illustrating the universality of the underlying systematics.

\section{SU(2) gauge theory}
\label{SU2sec}
The partition function of 4d $SU(2)$ gauge theory in the $\Omega$-background, parameterized by $\ep_1$ and $\ep_2$,  and denoted as $Z_\Omega$, is a non-trivial function over Coloumb moduli space. As is well known, there are three interesting points in moduli space. Namely, the weak coupling regime with massless gauge bosons (vector-multiplets), and two strongly coupled regimes, with either a massless dyon or monopole (hyper-multiplets). The two strongly coupled regimes are related by a $\Z_2$ symmetry, and hence can be treated simultaneously.

The expansion of the partition function near one of these points in moduli space can be split into two parts, \ie,
\beq\eqlabel{ZomegaDef}
Z_\Omega = Z_{sing} \times Z_{reg}\,.
\eq
$Z_{sing}$ refers to the contribution from massless vector, respectively hypermultiplets, while $Z_{reg}$ denotes the part of the partition function regular in, both, the dynamical scale $\Lambda$ and the Coulomb-parameter (flat coordinate) at the point of expansion. 

Of particular interest for us here is the so-called Nekrasov-Shatashvili limit of the partition function, defined as 
\beq\eqlabel{DefNSlimit}
\Fcal:= \lim_{\ep_2\rightarrow 0} \ep_1\ep_2\log Z_\Omega\,.
\eq
(Note that our definition is slightly different than the original one of \cite{NS09}, \ie, we include an overall $\ep_1$.) Clearly, according to \req{ZomegaDef} $\Fcal$ can be split as $\Fcal= \Fcal_{sing}+\Fcal_{reg}$. We will relabel the parameter $\ep_1$ surviving the limit as $\hbar$. In order to make contact with the limits taken in the previous $\beta$-ensemble section, one should note that under the redefinition $\ep_1=g_s, \ep_2=\beta g_s$, \req{DefNSlimit} is formally equivalent to \req{FQensembleDef} and \req{FNSensembleDef}, up to an irrelevant overall $g_s^2$.

The limit \req{DefNSlimit} has to be supplemented by the quantization condition \cite{NS09}
\beq\eqlabel{NSquantCond}
e^{\frac{1}{\hbar} \frac{\partial \Fcal(t)}{\partial t}}=1\,,
\eq
where $t$ is a flat coordinate near the point of expansion in the Coulomb moduli space. The condition \req{NSquantCond} ensures that we sit in a supersymmetric vacua of the effective 2d theory \cite{NS09}. One should keep in mind that the NS free energy $\Fcal$ defined above is a function of $\hbar$. In particular, at $\hbar=0$ one recovers the well-known Seiberg-Witten prepotential $\Fcal_{SW}$, \ie, $\Fcal(\hbar=0)=\Fcal_{SW}$ \cite{N02}. 

According to Seiberg and Witten, the prepotential $\Fcal_{SW}$ is fully encoded in a curve $\Sigma$ equipped with a meromorphic differential $\lambda_{SW}$. In detail, from the period integrals over the A- and B-cycle of the curve, $a=\oint_{\mathcal A} \lambda_{SW}$ and $a_D=\oint_{\mathcal B}\lambda_{SW}$, $\Fcal_{SW}$ is recovered via the special geometry (or integrability) relation
\beq\eqlabel{SpecialGeoRelation}
\left(
\begin{matrix}
a_D\\
a
\end{matrix}
\right)
=
\left(
\begin{matrix}
\frac{\partial\Fcal_{SW}(a)}{\partial a}\\
\frac{\partial\Fcal_{SW}(a_D)}{\partial a_D}\\
\end{matrix}
\right)\,,
\eq
where the top row holds at weak coupling, and the bottom row at strong coupling. The preferred flat coordinate $t$ is denoted at weak coupling by $t=a$ and at strong coupling by $t=a_D$. It is convenient to define $\Pi(t):=\frac{\partial\Fcal(t)}{\partial t}$. Note that $\Pi$ is either a series in $\Lambda$ or in $1/\Lambda$. In terms of $\Pi$, the condition $\req{NSquantCond}$ reads $e^{\Pi/\hbar}=1$. 

One should note that the choice of pair $(\Sigma, \lambda_{SW})$ is not unique. Here, we stick with the original curve of \cite{SW94}, namely we take $\Sigma: y^2=(z^2-\Lambda^2)(z-u)$ with differential $\lambda_{SW}=\frac{\sqrt{2}}{2\pi} \frac{z-u}{y(z)}dz$. This choice of curve is convenient for our purposes, as under a change of coordinate $z\rightarrow \Lambda \cos x$ the differential translates to \cite{GKMMM95}
$$
\lambda_{SW}=\frac{1}{2\pi} \sqrt{2(u-\Lambda^2\cos x)}\,.
$$
Hence, the periods are just the classical action integrals $\frac{1}{2\pi}\oint dx\, p(x)$ of a Sine-Gordon model with momentum $\frac{p^2}{2}=u-\Lambda^2\cos x$\,. Canonical quantization, \ie, imposing that $[x,p]=\ii \hbar$, yields the Schr\"odinger equation
\beq\eqlabel{SU2QM}
\left(-\frac{\hbar^2}{2}\partial_x^2 -u+\Lambda^2\cos\left( x\right)\right)\Psi(x)=0\,.
\eq

Under quantization, the differential form $\lambda_{SW}$, and so the action integrals, become $\hbar$-dependent functions. The classical curve $\Sigma$ equipped with the $\hbar$-dependent differential will be referred to as quantum geometry, with quantum periods given by integrating the quantum differential over the classical cycles, as in section \ref{pertquantgeo}. The remarkable results of \cite{NS09} can be reformulated as the statement that the NS limit of the 4d gauge theory \req{DefNSlimit} is equivalent to the free energy computed from the quantum geometry via the special geometry relation \req{SpecialGeoRelation} for the quantum periods \cite{MM09}. 

A remark is in order. Different choices of pairs $(\Sigma,\lambda_{SW})$ may lead to different normalization schemes (see also the remark in \cite{SW94b}). In order to compare/translate results obtained for the choice of pair above (and hence for the Schr\"odinger equation \req{SU2QM}) to other results in the literature, a rescaling of variables might be needed. Here, we mainly compare to \cite{N02} for weak coupling and to \cite{KW10a} (with tree-level geometry borrowed from \cite{KLT95}) for strong coupling. In both cases we have to rescale
\beq\eqlabel{RescaleForSU2}
a\rightarrow 2a\,,\,\,\,\,\, \hbar\rightarrow 2\hbar\,.
\eq
The comparison with \cite{N02} needs as well an additional rescaling $\Lambda\rightarrow 2\Lambda$. One should note that \req{RescaleForSU2} implies that at strong coupling we have to send $\Fcal(a_D)\rightarrow 2\Fcal(a_D)$.

An important role will be played by the so-called quantum Matone relation \cite{M95,STY95,FFMP04}
\beq\eqlabel{qMatoneRelation}
u(t)=c_t\,\Lambda\frac{\partial\Fcal(t)}{\partial\Lambda} +c_{A} \hbar^2\,,
\eq
where $c_t$ and $c_A$ are constants whose precise value depends on the point of expansion in Coulomb moduli space. The precise value of $c_A$ is not of utmost importance for us, but if needed can be inferred from appendix \ref{appMasslessMultiplets} (see also \cite{H11} for a discussion of the anomaly term $\sim\hbar^2$). For our normalization conventions, we have at weak coupling $c_t=-1/2$ and at strong coupling $c_t=-1/4$. Taking the $\partial_t$ derivative of \req{qMatoneRelation} leads to the relation
\beq\eqlabel{DMatoneRel}
\boxed{
\frac{\partial u(t)}{\partial t} = c_t\, \Lambda\frac{\partial\Pi_\Lambda(t)}{\partial \Lambda}
}\,,
\eq
where $\Pi_\Lambda$ refers to the $\Lambda$-dependent part of $\Pi$. Matone's relation \req{qMatoneRelation} and its derivative \req{DMatoneRel} allow us to infer the $\Lambda$-dependent part of the perturbative quantum prepotential, denoted as $\Fcal_\Lambda$, respectively $\Pi_\Lambda$, from knowledge of the quantum mirror map $u(t)$. 

It will be useful to view \req{SU2QM} as a transformation of the canonical form of Mathieu's equation
\beq\eqlabel{MathieuEQ}
\left(\partial_x^2+\alpha-2q \cos2 x \right)\Psi(x)=0\,,
\eq
where $\alpha$ is usually referred to as characteristic number. In detail, substituting $x\rightarrow x/2$ and setting
\beq\eqlabel{MAtoSU2QM}
\alpha= \frac{8u}{\hbar^2}\,,\,\,\,\,\,q= \frac{4\Lambda^2}{\hbar^2}\,,
\eq
transforms Mathieu's equation \req{MathieuEQ} to \req{SU2QM}. The relation to Mathieu's equation has been advocated and made use of earlier in \cite{H10} (and followup works thereof). Its main utilization is to infer the perturbative expansion of (the $\hbar$-dependent) $u$. In detail, there are known closed recursive expressions for perturbative expansions of $\alpha$ in, both, $q$ small and large \cite{FP01}, and so for $u$, see appendix \ref{charNumbersExp}. 

There exists another redefinition which is of interest. Namely, via redefining $x\rightarrow 4x+\pi$, and setting
\beq\eqlabel{SU2QMtoCosineQM}
u=\hbar\Lambda \,\cale-\Lambda^2\,,\,\,\,\,\, \hbar = 16\hbar\,,
\eq 
the Schr\"odinger equation \req{SU2QM} turns into
\beq\eqlabel{CosineQM}
\left(-\frac{\hbar^2}{2}\partial_x^2 +\frac{\Lambda^2}{16}\left(1-\cos\left( 4x\right)\right)\right)\Psi(x)=\hbar\Lambda \,\cale\,\Psi(x)\,.
\eq
This Schr\"odinger equation is known as critical Sine-Gordon model (under substituting $\cos(4x)=1-2\sin^2(2x)$), and has been investigated for $\Lambda=1$ extensively from a non-perturbative point of view \cite{ZJJ04a,ZJJ04b} (and more recently in \cite{DU13,DU14}). Keeping $\Lambda$ general will simplify things (similar as the additional parameter $\delta$ introduced for the double-well potential in \cite{K13}, see \req{anHoscQMSE}). It will turn out to be useful to define a new coordinate $\t u$ as 
\beq\eqlabel{upDef}
\t u:= u+\Lambda^2\,.
\eq
In particular, we then have that the transition to the Sine-Gordon model \req{CosineQM} is given by (up to some irrelevant rescaling of $\hbar$)
\beq\eqlabel{tuQuantization}
\boxed{
\t u= \hbar\Lambda\, \cale
}\,,
\eq
\ie, a quantization of the complex structure parameter.

The upshot is that the NS limit of $SU(2)$ gauge theory is equivalent to the quantum Sine-Gordon model \req{CosineQM}, under the quantization \req{tuQuantization}. Since the derivation of \req{SU2QM} and the mapping to \req{CosineQM} is independent of any perturbative expansion, the non-perturbative completion of the NS limit of $SU(2)$ gauge theory at the point \req{tuQuantization} in complex structure moduli space can be inferred from well established quantum mechanical results. We have however to make one assumption. Namely that the Matone relation \req{qMatoneRelation} holds as well non-perturbatively. We can offer at least one hint that this is indeed the case. Conceptually, the Matone relation is nothing else than the relation \req{fQMDef} observed before for the $\beta$-ensemble. As the relation for the $\beta$-ensemble holds non-perturbatively, it is expectable that so does \req{qMatoneRelation}. Phrased differently, the core assumption we make is that the non-perturbative corrections to the free energy are generally linked to non-perturbative corrections to the complex structure moduli (mirror map) of the underlying quantum geometry, if classically relations like \req{fQMDef} (of which \req{qMatoneRelation} is a special case) exists, \ie, we assume that section \ref{EnsembleNPcompletion} describes general facts.

As at the point \req{tuQuantization} in moduli space the NS limit of $SU(2)$ gauge theory is described by the Sine-Gordon model \req{CosineQM}, let us first discuss the non-perturbative expansion/completion of \req{CosineQM} in some more detail. As we will see later, this is in fact the case of relevance in general.

\section{Quantum Sine-Gordon}
\label{SGmodel}
\subsection{The model}
\label{SGintro}

The potential in the Sine-Gordon Schr\"odinger equation \req{CosineQM} is periodic under $\Tcal:x\rightarrow x+\pi/2$ (with minimal period). Due to Floquet's theorem there exists a particular solution $\Psi_\phi$ of the form
\beq\eqlabel{FloquetSolution}
\Psi_\phi(x)=e^{\\i \phi x}\, p(x)\,,
\eq
with $p(x)$ periodic under $\Tcal$ and $\phi$ constant, usually referred to as characteristic exponent. Clearly, the solution \req{FloquetSolution} picks up a phase under action of $\Tcal$, \ie,
$$
\Tcal\Psi_{\phi}(x)=e^{\ii \theta}\,\Psi_{\phi}(x)\,,
$$
with $\theta=\ii\pi \phi/2$. 

The exact quantization condition has been conjectured to be given by \cite{ZJJ04a,ZJJ04b}
\beq\eqlabel{CosineExactQuantCond}
\left(\frac{2}{\hbar}\right)^{-B(\cale)} \frac{e^{A(\cale)/2}}{\Gamma(\frac{1}{2}-B(\cale))}+\left(-\frac{2}{\hbar}\right)^{B(\cale)} \frac{e^{-A(\cale)/2}}{\Gamma(\frac{1}{2}+B(\cale))}=\sqrt{\frac{2}{\pi}}\,\cos \theta\,,
\eq
with generating functions $A(\cale)$ and $B(\cale)$. The function $A(\cale)$ takes the form 
\beq\eqlabel{Asplit}
A(\cale)=\frac{\Lambda}{\hbar}+A_p(\cale)\,,
\eq
with $A_p(\cale)$ non-singular as $\hbar\rightarrow 0$. This suggests to introduce an instanton counting parameter 
$$
\xi:= e^{-\frac{\Lambda}{2\hbar}}\,.
$$
(The reason for the factor of $1/2$ lies in the overall factor of $A$ in \req{CosineExactQuantCond}).
Accordingly, we expand the energy $\cale$ into a series in $\xi$. The terms of order $\xi^0$ will be referred to as perturbative energy $\cale_p$, while higher order terms in $\xi$ will be referred to as non-perturbative energy $\cale_{np}$, \ie, we split $\cale=\cale_p+\cale_{np}$, such that
$$
\cale_{np}=\sum_{n=1}^\infty E^{(n)}_{np} \, \xi^n\,,
$$
and refer to $E^{(n)}_{np}$ as the $n$-instanton non-perturbative energy. 

Using \req{Asplit}, and with help of Euler's reflection formula 
\beq\eqlabel{EulerFormula}
\Gamma(1-z)\Gamma(z)=\frac{\pi}{\sin(\pi z)}\,,
\eq
we rewrite the exact quantization condition \req{CosineExactQuantCond} as
\beq\eqlabel{CosineExactQuantCond2}
\frac{\cos(\pi B(\cale))}{\pi}=\left(\frac{2}{\hbar}\right)^{B(\cale)}\frac{\sqrt{2}e^{-A_p(\cale)/2}\cos\theta }{\sqrt{\pi}\,\Gamma(\frac{1}{2}+B(\cale))}\,\xi-(-1)^{B(\cale)}\left(\frac{2}{\hbar}\right)^{2B(\cale)} \frac{e^{-A_p(\cale)}}{\Gamma(\frac{1}{2}+B(\cale))^2}\,\xi^2
\,.
\eq
Clearly, we recover at order $\xi^0$, the usual perturbative Bohr-Sommerfeld quantization condition 
\beq\eqlabel{CosineQMPertQuantCond}
B(\cale_p)=N+1/2\,,
\eq
with $N$ integer. Hence, the generating function $B(\cale_p)$ is essentially the inverse of $\cale_p(N)$. The latter is easily obtainable via noting that via redefining $x\rightarrow 2x+\pi/2$ and setting
\beq\eqlabel{MAtoCosineQM}
\alpha=\frac{1}{2\hbar^2}\left(\hbar\Lambda \,\cale-\frac{\Lambda^2}{16}\right)\,,\,\,\,\,\,q=\frac{\Lambda^2}{64\hbar^2}\,,
\eq
in Mathieu's equation \req{MathieuEQ}, yields the Sine-Gordon model \req{CosineQM}. Hence, from the known perturbative expansion of $\alpha$ we can immediately read of $\cale_p$ (see appendix \ref{charNumbersExp}).

It has been observed in \cite{DU13,DU14}, that the functions $A(\cale_p)$ and $B(\cale_p)$ are not independent, but that there exists a derivative relation between them. Here, similar as in \cite{K13}, we observe, due to the introduction of the additional parameter $\Lambda$, a different, but related derivative relation. Namely, one can check order by order in $\hbar$ that via integrating
\beq\eqlabel{EArelation}
\boxed{
\frac{\partial\cale_p(N)}{\partial N}=\hbar\,\frac{\partial A(N)}{\partial\Lambda}
}\,,
\eq
one can recover at $\Lambda=1$ the known $A(N)$ (\cf, \req{EpQM} and \req{ACosineQM}). A priori, the relation \req{EArelation} only determines $A(N)$ up to some integration constant, which however vanishes, as one can check.

For later comparison it will be useful to rewrite \req{CosineExactQuantCond} as follows. We define
\beq\eqlabel{BEQMsplit}
\Pi(\cale):=\Pi^+(\cale)+\Pi^-(\cale)+\ii\pi\,,
\eq
with
\beq\eqlabel{PipmBEdef}
\Pi^\pm(\cale):=\pm \log\Gamma\left(\frac{1}{2}\pm B(\cale)\right)-B(\cale)\log\left(\mp\frac{2}{\hbar}\right)+\frac{1}{2}A(\cale)\,.
\eq
Under these definitions the exact quantization condition \req{CosineExactQuantCond} reads
\beq\eqlabel{PIexactQwTheta}
e^{\Pi^-(\cale)}+e^{-\Pi^+(\cale)}=\sqrt{\frac{2}{\pi}}\cos\theta\,.
\eq
In particular, for $\theta=\frac{\pi k}{2}$ with $k$ odd we simply have
\beq
e^{\Pi(\cale)}=1\,,
\eq
corresponding to 
\beq\eqlabel{SGspecialEQ}
\frac{\Gamma\left(\frac{1}{2}+B(\cale)\right)}{\Gamma\left(\frac{1}{2}-B(\cale)\right)}= -\left(\ii \frac{2}{\hbar}\right)^{2B(\cale)} e^{-A(\cale)}\,.
\eq
\subsection{Solving the exact quantization condition}
\label{SGEQsolution}
Following \cite{K13}, the exact quantization condition \req{CosineExactQuantCond2} can be easily solved order by order in $\xi$ via expanding $A(\cale)$ and $B(\cale)$ into powers of $\xi$.

\paragraph{1-instanton} 
We have
$$
B(\cale_p+\cale_{np})=N+1/2+ \frac{\partial B(\cale_p)}{\partial\cale_p} E_{np}^{(1)}\,\xi +\Ocal(\xi^2)\,.
$$
Hence, we learn that
\beq\eqlabel{ENPinst1}
\boxed{
E^{(1)}_{np}=\frac{\ii\sqrt{2}\cos\theta}{\sqrt{\pi}\,N!} \left(-\frac{2}{\hbar}\right)^{N+1/2}e^{-A_p(\cale_p)/2}\,\frac{\partial \cale_p}{\partial N}
}\,.
\eq
With the explicit expansion of $\cale_p(N)$ and $A(N)$ in powers of $\hbar$ (see \req{EpQM} and \req{ACosineQM}), we obtain for the first order in $\hbar$
$$
E_{np}^{(1)}(N,\theta)=\frac{\ii\sqrt{2}\cos\theta}{\sqrt{\pi}\,N!}\left(-\frac{2\Lambda}{\hbar}\right)^{N+1/2} \left(1-\frac{1}{4}\left(7+14N+6N^2\right) \frac{\hbar}{\Lambda} +\Ocal\left(\frac{\hbar^2}{\Lambda^2}\right)\right)\,.
$$

\paragraph{2-instanton}
We have
$$
B(\cale_p+\cale_{np})|_{\xi^2}=E^{(2)}_{np}\frac{\partial B(\cale_p)}{\partial \cale_p}+\frac{1}{2}\left(E_{np}^{(1)}\right)^2\frac{\partial^2 B(\cale_p)}{\partial^2\cale_p}.
$$
We further need the expansions
\beq
\begin{split}
e^{-A_p(\cale)/2}&=e^{-A_p(\cale_p)/2}\left(1-\frac{1}{2}\frac{\partial A_p(\cale_p)}{\partial\cale_p} E_{np}^{(1)}\,\xi+\Ocal(\xi^2)\right)\\
\frac{1}{\Gamma(\frac{1}{2}+B(\cale))}&=\frac{1}{N!}\left(1-\frac{\partial B(\cale_p)}{\partial\cale_p}\psi(1+N)E_{np}^{(1)}\,\xi+\Ocal(\xi^2)\right)\,,
\end{split}
\eq
with $\psi(z)$ the digamma function, and
$$
\left(\frac{2}{\hbar}\right)^{B(\cale)}=\left(\frac{2}{\hbar}\right)^{B(\cale_p)}\left(1+\log\left(\frac{2}{\hbar}\right)\frac{\partial B(\cale_p)}{\partial \cale_p} E_{np}^{(1)}\,\xi+\Ocal(\xi^2)\right)\,,
$$
in order to evaluate the second term in the right hand side of \req{CosineExactQuantCond2} up to order $\xi^2$. We obtain from this term a contribution of (making use of \req{ENPinst1}) 
$$
\sqrt{\frac{2}{\pi}}\frac{e^{-A_p(\cale_p)/2}\cos\theta}{N!}\left(\frac{2}{\hbar}\right)^{N+1/2} E_{np}^{(1)}\left(\left(\log\left(\frac{2}{\hbar}\right)-\psi(1+N)\right)\frac{\partial B(\cale_p)}{\partial \cale_p}-\frac{1}{2}\frac{\partial A_p(\cale_p)}{\partial\cale_p}  \right)\,.
$$
Taking the remaining terms into account, one infers 
\beq
\begin{split}
E_{np}^{(2)}=&-\frac{1}{2}\left(E_{np}^{(1)}\right)^2\frac{\partial^2 B(\cale_p)}{\partial^2\cale_p}\frac{\partial\cale_p}{\partial N}+ \ii^2\left(-\frac{2}{\hbar}\right)^{2N+1} \frac{e^{-A_p(\cale_p)}}{(N!)^2}\frac{\partial \cale_p}{\partial N}\\&+\left(E_{np}^{(1)}\right)^2 \left(\left(\log\left(\frac{2}{\hbar}\right)-\psi(1+N)\right)\frac{\partial B(\cale_p)}{\partial \cale_p}-\frac{1}{2}\frac{\partial A_p(\cale_p)}{\partial \cale_p}  \right)
\end{split}
\eq
(where we used again \req{ENPinst1} for simplification.)
This can be further simplified to 
\beq\eqlabel{E2inst}
\boxed{
\begin{split}
E_{np}^{(2)}=&\left(E_{np}^{(1)}\right)^2\left(\left(\log\left(\frac{2}{\hbar}\right)-\psi(1+N)+ \frac{ \pi}{2 \cos^2\theta}\right)\frac{\partial B(\cale_p)}{\partial \cale_p}\right.\\&\left.-\frac{1}{2}\left(\frac{\partial A_p(\cale_p)}{\partial \cale_p}+\frac{\partial\cale_p}{\partial N}\frac{\partial^2 B(\cale_p)}{\partial^2\cale_p} \right) \right)\,.
\end{split}
}\,,
\eq
We obtain for the first orders in $\hbar$
\beq
\begin{split}
E_{np}^{(2)}=\frac{2\cos^2\theta}{\pi\,(N!)^2}\left(\frac{2\Lambda}{\hbar}\right)^{2N+1}&\left(\left(\log\left(\frac{2}{\hbar}\right)-\psi(1+N)+ \frac{ \pi}{2 \cos^2\theta}\right)\right.\\
&\left(1-\frac{1}{2}\left(5+10N+6N^2\right)\frac{\hbar}{\Lambda} +\Ocal\left(\frac{\hbar^2}{\Lambda^2}\right)\right)\\
&\left.-\frac{1}{2}\left(5+6N\right)\hbar-\frac{1}{2}\left(18+41N+27N^2\right)\frac{\hbar^2}{\Lambda^2}+\Ocal\left(\frac{\hbar^3}{\Lambda^3}\right)\right)\,.
\end{split}
\eq
Higher order non-perturbative energies $E^{(n)}_{np}$ can be inferred in a similar fashion via expansion of \req{CosineExactQuantCond} into higher powers of $\xi$. Hence, the full (perturbative+non-perturbative) energy $\cale$ can be solved for analytically, order by order in the two expansion parameters $\hbar$ and $\xi$.

\subsection{Free energy}
\label{SGfenergySec}
Let us define a free energy $\Fcal_{SG}(N)$ via 
\beq\eqlabel{FviaE}
\boxed{
\cale(N)=:\hbar \frac{\partial}{\partial\Lambda} \Fcal_{SG}(N)
}\,.
\eq
Relation \req{EArelation} implies that we have sort of a special geometry relation 
\beq\eqlabel{FSGspecialGeoRel}
\Fcal_{SG}(N)=\int dN A(N)\,,
\eq
up to some integration constant $c$ (more precisely function). As $\cale$ and $A$ possess an expansion in $\xi$, so does $\Fcal_{SG}(N)$, \ie,
$$
\Fcal_{SG}(N)=\Fcal_p(N)+\sum_{n=1}^\infty \Fcal^{(n)}_{np}(N)\,\xi^n\,.
$$
We will refer to $\Fcal_{np}^{(n)}$ as the $n$-instanton free energy. The integration constant $c$ is also expanded into powers of $\xi$, with expansion coefficients $c^{(n)}$.

\paragraph{1-instanton}

With help of relation \req{EArelation} we can rewrite the 1-instanton energy as
$$
E_{np}^{(1)}\,\xi =-2\hbar\frac{\ii\sqrt{2}\cos\theta}{\sqrt{\pi}\,N!} \left(-\frac{2}{\hbar}\right)^{N+1/2}\frac{\partial}{\partial\Lambda}e^{-A(\cale_p)/2}\,.
$$
Then, 
$$
E_{np}^{(1)}\,\xi =-2\hbar\frac{\ii\sqrt{2}\cos\theta}{\sqrt{\pi}} \frac{\partial}{\partial\Lambda}e^{-\Pi^+(N)}\,.
$$
Due to \req{FviaE}, the 1-instanton non-perturbative free energy is obtained via integration over $\Lambda$, \ie,
$$
\Fcal_{np}^{(1)} := \frac{1}{\hbar}\int d\Lambda \,E^{(1)}_{np}+c^{(1)}\,.
$$
Clearly,
$$
\boxed{
\Fcal^{(1)}_{np}\,\xi=-2 \frac{\ii\sqrt{2}\cos\theta}{\sqrt{\pi}}\, e^{-\Pi^+(N)}+  c^{(1)}
}\,,
$$

\paragraph{2-instantons}
The 2-instanton energy takes the qualitative form, \cf, \req{E2inst} (see also \cite{K13})
$$
E^{(2)}_{np}=\left(E^{(1)}_{np}\right)^2 \left(C(N)\frac{\partial B(\cale_p)}{\partial\cale_p}-\frac{1}{2}\left(\frac{\partial A_p(\cale_p)}{\partial \cale_p}+\frac{\partial\cale_p}{\partial N}\frac{\partial^2 B(\cale_p)}{\partial^2\cale_p}\right)\right)\,,
$$
with $C(N)$ some constant (function of $N$). The first part involving $C(N)$ can be easily integrated over $\Lambda$, similar as for the 1-instanton energy. However, the latter term requires first some rewriting. We write
$$
\frac{\partial A_p(\cale_p)}{\partial \cale_p}+\frac{\partial\cale_p}{\partial N}\frac{\partial^2 B(\cale_p)}{\partial^2\cale_p}=\frac{\partial}{\partial\cale_p}\left(A(\cale_p)+\log \frac{\partial B(\cale_p)}{\partial \cale_p}\right)=\frac{\partial}{\partial\cale_p}\left(A(\cale_p)-\log \frac{\partial A(N)}{\partial \Lambda}\right)\,,
$$
under usage of \req{EArelation}, and note that
$$
\left(\frac{\partial A}{\partial\Lambda}\right)\frac{\partial}{\partial N}\left(A(\cale_p)-\log \frac{\partial A(N)}{\partial \Lambda}\right)=\left(\frac{\partial A(N)}{\partial\Lambda}\right)\left(\frac{\partial A(N)}{\partial N}\right)-\frac{\partial^2 A(N)}{\partial N\partial\Lambda}\,.
$$
Using the above two relations, leads us to 
\beq
\begin{split}
E^{(2)}_{np}\,\xi &= \hbar\frac{2\cos^2\theta}{\pi}\left(C(N)+\psi(1+N)-\log\left(-\frac{2}{\hbar}\right)-\frac{1}{2}\frac{\partial}{\partial N} \right)\frac{\partial}{\partial\Lambda}e^{-2\Pi^+(N)}\\
&=\hbar\frac{2\cos^2\theta}{\pi}\left( \frac{ \pi}{2 \cos^2\theta}-\frac{1}{2}\frac{\partial}{\partial N} \right)\frac{\partial}{\partial\Lambda}e^{-2\Pi^+(N)}\,.
\end{split}
\eq
We conclude that the 2-instanton free energy reads
$$
\boxed{
\Fcal^{(2)}_{np}\,\xi^2 = \left(1 +\frac{\cos^2\theta}{\pi}\frac{\partial}{\partial N} \right)e^{-2\Pi^+(N)}+c^{(2)}
}\,.
$$

Note that for the special values $\theta=\frac{\pi k}{2}$ with $k$ odd, introduced at the end of section \ref{SGintro}, we have that $\Fcal^{(1)}_{np}=0$ and
\beq\eqlabel{SU2F2Theta0}
\Fcal^{(2)}_{np}\, \xi^2=\, e^{-2\Pi^+(N)}+c^{(2)}\,.
\eq
This suggests the redefinition $\hat \xi= \xi^{2}$, such that $\Fcal_{np}^{(2)}\xi^2\rightarrow \hat \Fcal_{np}^{(1)} \hat \xi$, \ie, \req{SU2F2Theta0} is a 1-instanton correction in terms of $\hat \xi$.

\section{NP corrections to SU(2) gauge theory}
\label{NPsu2sec}
\subsection{Strong coupling}
Let us first discuss the strong coupling regime, \ie, the expansion near the monopole or dyon point. $\Fcal_{sing}$ is given by a contribution of a hyper-multiplet (two scalars) and can be deduced to be (\cf, \cite{GNY10})
$$
\Fcal_{sing}= \delta_{\hbar/2}(a_D,\Lambda)+\delta_{\hbar/2}(-a_D,\Lambda)\,,
$$
with function $\delta_{\ep}$ detailed in \req{Deltaep}. The period $a_D$ is the preferred flat coordinate near the strongly coupled regime. The regular part, $\Fcal_{reg}$, can for instance be inferred via analytic continuation from weak coupling, making use of the underlying special geometry and holomorphic anomaly equations (originating from modularity of the partition function) \cite{KW10a,KW10b}, as a series in $\hbar$ and $1/\Lambda$. However, it is more convenient to use the correspondence to the quantum system \req{SU2QM} and the quantum Matone relation \req{qMatoneRelation} to directly infer $\Fcal_{reg}$, as demonstrated in appendix \ref{charNumbersExp}.

The quantization condition of Nekrasov and Shatashvili, eq. \req{NSquantCond}, reads under making use of \req{Ddeltae}
\beq\eqlabel{SU2exactQuant}
\frac{\Gamma\left(\frac{1}{2}+\frac{2a_D}{\hbar}\right)}{\Gamma\left(\frac{1}{2}-\frac{2a_D}{\hbar}\right)}=\left(\frac{2}{\hbar}\right)^{4a_D/\hbar}e^{2\Pi_\Lambda(a_D)/\hbar}\,.
\eq
Note that the exact quantization condition \req{SGspecialEQ} of the Sine-Gordon model at $\theta=\frac{\pi k}{2}$ with $k$ odd, discussed in the previous section, can be recovered from \req{SU2exactQuant} via the substitutions
\beq\eqlabel{aDPiLtoSG}
a_D = \frac{\hbar}{2} B\,,\,\,\,\,\, \Pi_\Lambda = -\frac{\hbar}{2} A\,.
\eq 
(With a rescaling $\hbar\rightarrow 16\hbar$, \cf, \req{SU2QMtoCosineQM}, in $a_D$ and $\Pi_\Lambda$). It can be explicitly verified that these two relations indeed hold on a perturbative level, see \req{aDB} and \req{SU2strongAviaP}. The quantum version of the special geometry relation \req{SpecialGeoRelation} and the definition of $\Fcal_{SG}$ \req{FSGspecialGeoRel} imply that one has under the mapping \req{aDPiLtoSG},
\beq\eqlabel{FFSQrescale}
\Fcal\rightarrow -\frac{\hbar^2}{4}\Fcal_{SG}\,.
\eq
The derivative of the Matone relation \req{DMatoneRel} translates under \req{aDPiLtoSG}, making use of \req{tuQuantization}, to \req{EArelation}. Strictly speaking, via the substitutions \req{aDPiLtoSG} we recover \req{SGspecialEQ} only up to a relative phase. Hence, we expect that there will be also an overall phase-difference in the non-perturbative corrections. The reason for that can be found in the phase shift of the potential under going from \req{SU2QM} to \req{CosineQM}.

A remark is in order. The relation to the Sine-Gordon model tells us that \req{NSquantCond}, and so \req{SU2exactQuant}, is not the most general quantization condition, but rather \req{PIexactQwTheta} (under a suitable change of parameters). Hence, one should in fact introduce a theta angle in the effective 2d theory. For simplicity, we however consider here only the non-generic case given by \req{SU2exactQuant}.

It is important to note that $\Pi_\Lambda$ (expanded at strong coupling) possesses a constant term $-8$ of order $\Lambda^1$ and $\hbar^0$, \cf, \req{PiSU2Lambda}, leading to a factor of 
$$
\bar\xi:=e^{-\frac{16\Lambda}{\hbar}}\,,
$$
on the right-hand side of \req{SU2exactQuant} (with $\bar\xi\rightarrow \hat\xi$ under \req{aDPiLtoSG}). Since a priori there is no other term like this in \req{SU2exactQuant}, the relation \req{SU2exactQuant} can only be fulfilled if $a_D(\t u)$ receives corrections in powers of $\bar\xi$ (we use here the natural coordinate $\t u$ near the strongly coupled regime defined in \req{upDef}).

We take
\beq\eqlabel{tuNP}
\t u=\t u_p+\t u_{np}\,,
\eq
with 
$$
\t u_{np}=\sum_{n=1}^\infty \t u_{np}^{(n)}\,\bar\xi^n\,,
$$
similar as for the energy in section \ref{SGEQsolution}. Expansion of $a_D(\t u_p+\t u_{np})$ in $\bar\xi$ then yields
\beq\eqlabel{aDuExpansion}
a_D(\t u)=a_D(\t u_p)+\frac{\partial a_D(\t u_p)}{\partial \t u_p} \t u_{np}^{(1)}\,\bar\xi+ \Ocal\left(\bar\xi^2\right)\,.
\eq
Inserting \req{aDuExpansion} into \req{SU2exactQuant}, we infer from the order $\bar\xi^0$ that
\beq\eqlabel{SU2pertQuantCond}
a_D(\t u_p) = \frac{1}{2}\left(N+\frac{1}{2}\right)\,\hbar\,,
\eq 
with $N$ integer, must hold. We will refer to \req{SU2pertQuantCond} as perturbative quantization condition. Note that \req{SU2pertQuantCond} is equivalent to \req{aDPiLtoSG}, as $B(\cale_p)=N+1/2$. Hence, the order $\bar\xi^0$ enforces a transition to the Sine-Gordon model, parameterized as in \req{CosineQM}. The upshot is, that we could read of the non-perturbative completion of $\Fcal$ at strong coupling directly from sections \ref{SGEQsolution} and \ref{SGfenergySec} (up to some overall phase). For illustration, let us however give some more details below. 

\paragraph{1-instanton}
As in section \ref{SGintro}, it is convenient to rewrite the exact quantization condition via Euler's reflection formula \req{EulerFormula} as
$$
\frac{\cos\left(\frac{2 \pi a_D(\t u)}{\hbar}\right)}{\pi}=\left(\frac{2}{\hbar}\right)^{4a_D(\t u)/\hbar}\frac{e^{2\Pi_\Lambda(a_D(\t u))/\hbar}}{\Gamma^2\left(\frac{1}{2}+\frac{2 a_D(\t u)}{\hbar}\right)}\,,
$$
Inserting the expansion \req{aDuExpansion} then yields at order $\bar\xi^1$
\beq\eqlabel{SU2uNP1}
\t u_{np}^{(1)}(a_D(\t u_p))\,\bar\xi = -\frac{\hbar}{2\sin\left(\frac{2\pi a_D(\t u_p)}{\hbar}\right)} \left(\frac{2}{\hbar}\right)^{4a_D(\t u_p)/\hbar} \frac{e^{2\Pi_\Lambda(a_D(\t u_p))/\hbar}}{\Gamma^2\left(\frac{1}{2}+\frac{2 a_D(\t u_p)}{\hbar}\right)}\frac{\partial \t u_p}{\partial a_D}\,.
\eq
Similar as we did for the Sine-Gordon model, it is useful to split the period $\Pi$ as $\Pi=\Pi^+ + \Pi^-$
with (\cf, \req{Ddeltae})
\beq\eqlabel{PiStrongCouplingDef}
\begin{split}
\Pi^\pm(a_D):&=\frac{\partial \delta_{\hbar/2}(\pm a_D,\Lambda)}{\partial a_D}+\frac{1}{2}\frac{\partial\Fcal_{reg}}{\partial a_D}\\
&= \mp \frac{\hbar}{2}\log\Gamma\left(\frac{1}{2}\pm\frac{2a_D}{\hbar}\right)+a_D\log\left(\frac{2}{\hbar}\right)\pm\frac{\hbar}{4}\log 2\pi +\frac{1}{2} \Pi_\Lambda(a_D)\,.
\end{split}
\eq
Integrating \req{SU2uNP1} via \req{qMatoneRelation} and \req{DMatoneRel} leads to
$$
\boxed{
\Fcal^{(1)}_{np}\, \bar\xi=-\frac{\hbar^2}{8\pi\sin\left(\frac{2\pi a_D(\t u_p)}{\hbar}\right)}\,  e^{4\Pi^+(a_D(\t u_p))/\hbar}
}\,.
$$
However, we still have to impose the perturbative quantization condition \req{SU2pertQuantCond}. Since under \req{SU2pertQuantCond}
$$
\Pi^\pm(a_D(\t u_p))\rightarrow -\frac{\hbar}{2}\Pi^\pm(B(\cale_p))-\frac{\hbar}{2}B(\cale_p)\log\left(\mp 1\right)\pm \frac{\hbar}{4}\log 2\pi\,,
$$
with $\Pi^\pm(B(\cale_p))$ as defined in \req{PipmBEdef}, we have that the above expression for $\Fcal^{(1)}_{np}\bar\xi$ turns into $\hat \Fcal_{np}^{(1)} \hat\xi$ given in \req{SU2F2Theta0}, up to the rescaling \req{FFSQrescale} and some overall phase.

 Recall from section \ref{SGfenergySec} that under taking $\theta=\frac{\pi k}{2}$ with $k$ odd the sector $\xi^n$ with $n$ odd vanishes and one can map $\xi^{2n}\rightarrow \hat \xi^n$. Hence, it is more efficient to calculate the instanton corrections as above for $\theta=\frac{\pi k}{2}$, than for general $\theta$. It is straight-forward, but somewhat elaborative to calculate higher order instanton corrections, which we leave to the interested reader.

\subsection{Weak coupling}

For expansion at weak coupling, $\Fcal_{sing}$ is usually referred to as perturbative contribution and $\Fcal_{reg}$ as instanton part. In this case, $\Fcal_{sing}$ is given by a contribution of two vector-multiplets and reads \cite{NO03,NY03} (recall the rescaling \req{RescaleForSU2}) 
$$
\Fcal_{sing} = -\left(\gamma_{\hbar/2}(a,\Lambda)+\gamma_{\hbar/2}(-a,\Lambda)\right)\,,
$$
with function $\gamma_{\hbar}$ detailed in equation \req{gammaeInt} of appendix \ref{vectorContribution}, and $a$ being the Coulomb-parameter expanded near the weakly coupled regime in moduli space. The regular part, $\Fcal_{reg}$, is obtainable as a series in $\Lambda$ via localization, following \cite{N02}. Hence, with help of \req{Dgammae} we deduce that the condition \req{NSquantCond} reads in this case
\beq\eqlabel{SUweakCoupExactQCond}
\frac{\Gamma\left(1+\frac{2a}{\hbar}\right)}{\Gamma\left(1-\frac{2a}{\hbar}\right)} =\left(\frac{2\Lambda}{\hbar}\right)^{\frac{4a}{\hbar}} e^{-\frac{2}{\hbar}\frac{\partial \Fcal_{reg}}{\partial a}}\,.
\eq
(\cf, the similar expression previously derived in \cite{NS09}.) Qualitatively, the condition \req{SUweakCoupExactQCond} looks very similar to the previously considered exact quantization conditions \req{SGspecialEQ} and \req{SU2exactQuant}. However, thinks are quite different at weak coupling, as we will see below.

Making use of Euler's reflection formula \req{EulerFormula}, the condition \req{SUweakCoupExactQCond} turns into 
\beq\eqlabel{weakCexactQcond}
\frac{\sin\left(\frac{2\pi a}{\hbar}\right)}{\pi}=\frac{1}{\Gamma\left(1+\frac{2a}{\hbar}\right)\Gamma\left(\frac{2a}{\hbar}\right)}\left(\frac{2\Lambda}{\hbar}\right)^{\frac{4a}{\hbar}}\,e^{-\frac{2}{\hbar}\frac{\partial\Fcal_{reg}}{\partial a}(a)}\,.
\eq
The $\Lambda$-dependent part of the free energy at weak coupling reads (see appendix \ref{charNumbersExp})
$$
\Fcal_\Lambda(a)=-\left(a^2+\frac{\hbar^2}{24}\right)\log\Lambda+\Fcal_{reg}{(a)}\,.
$$
Correspondingly,
$$
\Pi_\Lambda(a)=-2a\log\Lambda+\frac{\partial \Fcal_{reg}(a)}{\partial a}\,.
$$
In contrast to the previous expansion at strong coupling, we have that $\frac{\partial\Fcal_{reg}}{\partial a}$ does not possess a distinguished constant term (\ie, $\sim a^0$) of order $\hbar^0$, \cf, \req{PiLweakExpansion}. Hence, it is not immediately clear what should be taken as instanton counting parameter. Therefore, we expand $\Pi_\Lambda$ as a series in $\hbar$, with expansion coefficients $\Pi^{(n)}_\Lambda$, and write
$$
\Pi_\Lambda(a)=\Pi^{(0)}_\Lambda(a)+\Pi_\Lambda^{\hbar}(a)\,,
$$
with $\Pi_\Lambda^{\hbar}=\sum_{n=1}^\infty \Pi^{(n)}_\Lambda(a)\, \hbar^n$. Similarly for $\Fcal_{reg}$. The expansion suggests to take 
$$
\boxed{
\t\xi:= e^{-\frac{2}{\hbar}\frac{\partial\Fcal^{(0)}_{reg}}{\partial a}(a_p)}
}\,,
$$
as instanton counting parameter. Then, as for strong coupling, since a priori there is no term of order $e^{\#/\hbar}$ on the left-hand side of \req{weakCexactQcond}, the relation \req{SUweakCoupExactQCond} can only be fulfilled if $a$ receives exponential corrections, \ie, 
\beq\eqlabel{aExpansion}
a=a_p+\sum_{n=1}^\infty a^{(n)}\,\t\xi^n\,.
\eq
Inserting \req{aExpansion} into $\frac{\partial\Fcal^{(0)}_{reg}}{\partial a}(a)$, transforms the exact quantization condition \req{weakCexactQcond} to
\beq\eqlabel{SU2weakFullQC}
\frac{\sin\left(\frac{2\pi a}{\hbar}\right)}{\pi}=\frac{e^{-\frac{2}{\hbar } \frac{\partial \Fcal_{reg}^{\hbar}}{\partial a}(a)   }}{\Gamma\left(1+\frac{2a}{\hbar}\right)\Gamma\left(\frac{2a}{\hbar}\right)}\left(\frac{2\Lambda}{\hbar}\right)^{\frac{4a}{\hbar}} \,\left(\t \xi-\frac{2a^{(1)}}{\hbar} \frac{\partial^2 \Fcal^{(0)}_{reg} \left(a_p\right)}{\partial^2 a_p}\, \t\xi^2+\Ocal\left(\t\xi^3\right)\right)\,.
\eq
The order $\t\xi^0$ of \req{SU2weakFullQC} yields the perturbative quantization condition
\beq\eqlabel{SU2weaka0N}
\boxed{
a_p=\frac{N}{2} \,\hbar
}
\,,
\eq
with $N$ integer. Note that small $\t\xi$ requires $\hbar\ll\Lambda$. However, imposing \req{SU2weaka0N} onto $\t \xi$ (which depends on $a_p$), we see that in fact $\t \xi$ is highly suppressed (\cf, \req{PiLweakExpansion}), essentially yielding $\t \xi =0$ such that non-perturbative corrections are negligible at weak coupling.  

\paragraph{Remark added}
For $\Lambda \ll \hbar$ the expansion parameter $\t\xi$ is not anymore small and it is more suitable to expand instead in terms of $\check\xi\sim \left(\frac{2 e}{\hbar N}\right)^{2N}$, \cf, \cite{BD15}. (Making use of the asymptotic expansion \req{LogGasymp} of the $\Gamma$-function for large $a/\hbar$.) In particular, in terms of $\check\xi$ we can parameterize the corrections to the free energy at weak coupling similar as in the strong coupling section. For instance, the first order correction can be parameterized in terms of $\check\xi$ analogously, \ie, 
$$
\Fcal^{(1)}_{np}\,\check\xi\sim e^{-2\Pi^+(N)}\,,
$$
with $\Pi^+$ similarly defined as in \req{PiStrongCouplingDef}.

\acknowledgments
D.K. likes to thank Seok Kim for useful discussions, and the Berkeley Center for Theoretical Physics and CERN for hospitality during part of this work. The work of D.K. has been supported by the National Research Foundation of Korea, Grant No. 2012R1A2A2A02046739.

\appendix

\section{Derivation of Schr\"odinger equation}
\label{SEQDerivation}

Acting with $\partial_x^2$ onto $\Psi_{k,h}(x)$, as defined in \req{Psidef}, yields the differential equation
\beq\eqlabel{Sdereq1}
\begin{split}
\Psi''_{k,h}(x)=&\,\frac{k^2}{g_s^2}\left(\left(W'(x)\right)^2- \frac{g_s}{k} W''(x) \right)\Psi_{k,h}(x)\\
&-\frac{2k}{g_s}W'(x)e^{-\frac{k}{g_s}W(x)}\frac{\partial_x \corr{.}_h}{\corr{.}}+e^{-\frac{k}{g_s}W(x)}\frac{\partial^2_x \corr{.}_h}{\corr{.}}.
\end{split}
\eq
On the one hand, acting with the derivatives on the correlators gives
\beq
\eqlabel{Sdereq2}
\partial_x\corr{.}_h=\sum_{i=1}^N\corr{\frac{h}{x-\lambda_i}}_h\,,\,\,\,\,\,\partial_x^2\corr{.}_h= \sum_{i\neq j}^N\corr{\frac{2h^2}{(\lambda_i-\lambda_j)(x-\lambda_i)}}_h+\sum_{i=1}^N\corr{\frac{h(h-1)}{(x-\lambda_i)^2}}_h\,.
\eq
While on the other hand the Ward identity
$$
0=\sum_{i=1}^N \int_\Ccal [d\lambda] \frac{\partial}{\partial \lambda_i}\left(\frac{1}{x-\lambda_i}\Delta(\lambda)^{2\beta}\, \psi_h(x) \,e^{-\frac{\beta}{g_s}\sum_{i=1}^N W(\lambda_i)} \right)\,,
$$
leads to the relation
\beq
\eqlabel{Sdereq3}
\sum_{i=1}^N\corr{\frac{h(h-1)}{(x-\lambda_i)^2}}_h=\frac{h\beta}{g_s}\sum_{i=1}^N\corr{\frac{W'(\lambda_i)}{x-\lambda_i}}_h-\sum_{i\neq j}^N\corr{\frac{2h\beta}{(\lambda_i-\lambda_j)(x-\lambda_i)}}_h\,.
\eq
For $h\neq 1$ we infer via substituting \req{Sdereq2} and \req{Sdereq3} into \req{Sdereq1} 
\beq
\begin{split}
\Psi''_{k,h}(x)=&\,\frac{k^2}{g_s^2}\left(\left(W'(x)\right)^2- \frac{g_s}{k} W''(x) \right)\Psi_{k,h}(x)\\
&-\frac{h}{g_s\corr{.}}e^{-\frac{k}{g_s}W(x)}\sum_{i=1}^N \corr{\frac{2k W'(x)-\beta W'(\lambda_i)}{x-\lambda_i}}_h\\
&+\frac{2h}{\corr{.}}e^{-\frac{k}{g_s}W(x)}\sum_{i\neq j}^N\corr{\frac{h-\beta}{(\lambda_i-\lambda_j)(x-\lambda_i)}}_h.
\end{split}
\eq
Similarly, we obtain for $h=1$ the differential equation
\beq
\Psi''_{k,1}(x)=\frac{k^2}{g_s^2}\left(\left(W'(x)\right)^2- \frac{g_s}{k} W''(x) \right)\Psi_{k,1}(x)-\frac{e^{-\frac{k}{g_s}W(x)}}{g_s\corr{.}}\sum_{i=1}^N \corr{\frac{2k W'(x)-W'(\lambda_i)}{x-\lambda_i}}_1\,.
\eq
Hence, we find that the two natural brane solutions \req{Psi12paras} lead to the differential equation \req{UnifiedSE}. The operator $\hat d(x)$, defined in \req{hfxDef}, can be commuted to the left in \req{UnifiedSE}, using the commutation relations
$$
\h D(x)\left(e^{-\frac{c_i}{2g_s}W(x)} \Ocal\right)= -\frac{c_i}{2g_s}\h D(x)\left( W(x)\right) e^{-\frac{c_i}{2g_s}W(x)}\Ocal+ e^{-\frac{c_i}{2g_s}W(x)} \h D(x)\Ocal\,,
$$
and
$$
\hat D(x) \left(\frac{1}{\corr{.}}\Ocal\right)=-\hat D(x)\left(\log\corr{.}\right)\frac{1}{\corr{.}}\Ocal+\frac{1}{\corr{.}}\hat D(x) \Ocal\,.
$$
We infer
$$
-\frac{c_i^2}{g_s\corr{.}}e^{-\frac{c_i}{2g_s}W(x)} \h f(x)\corr{.}_{c_i}=-\frac{c_i^2}{g_s} \left(c(x)-\frac{c_i}{2\beta} \h D(x)\left(W(x)+\frac{2g_s}{c_i}\Fcal_\Ccal\right)-\frac{g_s}{\beta}\h D(x)\right)\Psi_i(x)\,,
$$
with the free energy $\Fcal_\Ccal(N;\beta,g_s):=\log\corr{.}$. Hence, under an additional rescaling $g_s\rightarrow c_i\,g_s/2$ we obtain the multi-time dependent Schr\"odinger equation
\beq\eqlabel{tDepSE}
\begin{split}
g_s^2\,\Psi_i''(x)=&\,\left(\left(W'(x)\right)^2- g_s\left(W''(x)+2 c_i\, c(x)-\frac{c_i^2}{\beta}\h D(x)W(x) \right) \right)\Psi_{i}(x)\\
&-\frac{c_i^2 g_s^2}{\beta}\left( \h D(x)\left(\Fcal_\Ccal(N;\beta, c_i\,g_s/2)\right)-\h D(x)\right)\Psi_i(x)\,.
\end{split}
\eq
For $\Psi_1$ we can decouple the $\hat D$ operator via taking the limit $\beta\rightarrow 0$ (since $c_1=\beta$), leading to the time-independent Schr\"odinger equation \req{tiSEQgeneral}.

\section{Characteristic numbers}
\label{charNumbersExp}
It is well known that periodicity of the solutions of the Mathieu equation \req{MathieuEQ}, with different solutions of same periodicity parameterized by a parameter $r$ (integer for periodicity of $\pi$ or $2\pi$ and a rational fraction $r=n/s$ for periodicity of $2\pi s$), require that the characteristic number $\alpha$ and the parameters $q$ and $r$ are interrelated (\cf, \cite{ML51}), \ie, $\alpha$ is a function of $r$ and $q$. Convenient recursive formula for the $\alpha(r,q)$, expanded for either large or small $q$ have been obtained in \cite{FP01}. 

For small $q$ and rational $r$, $\alpha$ can be expanded as
\beq\eqlabel{alphaSmallq}
\alpha(r,q)=r^2+\sum_{k=1}^\infty \left(A_2^{2k-1}+A_{-2}^{2k-1}\right)\, q^{2k}\,,
\eq
with $A^{k}_{2i}$ recursively given by
\beq\eqlabel{Adef}
A^{k}_{2i}=-\frac{1}{4i(r+i)}\left(A^{k-1}_{2i+2}+A^{k-1}_{2i-2}-\sum_{j=1}^{\frac{k-|i|}{2}}\left(A_2^{2j-1}+A_{-2}^{2j-1}\right)A^{k-2j}_{2i}\right)\,,
\eq
with $A^0_0=1, A^k_0=0,A^k_{-2r}=0$ and $A^k_{2i}=0$ for $|i|>k$.

The asymptotic expansion of $\alpha(r,q)$ for large $q$ and integer $r$ reads
\beq\eqlabel{alphaExpansion}
\alpha(r,q)=-2q+2(2r+1)q^{1/2}-\frac{1}{2}\left(r^2+r+\frac{1}{2}\right)-\sum_{k=1}^\infty \gamma_{k}^{(r)}\, q^{-k/2}\,,
\eq
with
$$
\gamma_k^{(r)}=-\frac{1}{4^{k+1}}\left(G^k_{-4}+2G^{k}_{-2}-2(r+1)_2G^k_2+(r+1)_4G^k_4\right)\,,
$$
where $(m)_n$ refers to the Pochhammer symbol and $G^k_i$ is recursively defined via
\beq\eqlabel{Gdef}
\begin{split}
G^k_{2i}=&-\frac{1}{4i}\left(\frac{1}{2} G^{k-1}_{2i-4}+G^{k-1}_{2i-2}-2i(2r+2i+1)G^{k-1}_{2i}-(r+2i+1)_2G^{k-1}_{2i+2}\right.\\
&\left. +\frac{1}{2}(r+2i+1)_4G^{k-1}_{2i+4}+\frac{1}{2} \sum_{j=1}^{k-\left\lfloor \frac{|i|+1}{2} \right\rfloor}4^j\gamma_{j-1}^{(r)}G^{k-j}_{2i}\right)\,.
\end{split}
\eq
(Note that \req{Gdef} corrects a typo in the original equation of \cite{FP01}.) 

These recursive relations allow us to determine $\alpha(r,q)$ as a series in $q$ or $1/q$ rather efficiently.

\paragraph{Sine-Gordon}
The perturbative quantum mechanical energy $\cale_p(N)$ of the Sine-Gordon model \req{CosineQM} can be easily obtained from the characteristic numbers \req{alphaExpansion} via the change of variables \req{MAtoCosineQM}. However, for that, we first have to fix the relation between the parameter $r$ of $\alpha(r,q)$ and the energy level $N$. We know that
$$
\cale_p(N)=N+1/2+\Ocal\left(\hbar\right)\,.
$$ 
Hence, comparing via \req{MAtoCosineQM} with the second term of \req{alphaExpansion}, we deduce that actually 
$$
\boxed{
r=N
}\,,
$$
must hold. Hence, we obtain from \req{alphaExpansion} for the first few orders in $\hbar$ of the perturbative energy the expansion
\beq\eqlabel{EpQM}
\begin{split}
\cale_p(N)=&\, \frac{1}{2}(1+2N) -\frac{1}{2}(1+2N+2N^2)\frac{\hbar}{\Lambda}-\frac{1}{2}(1+3N+3N^2+2N^3) \left(\frac{\hbar}{\Lambda}\right)^2\\
&-\frac{1}{2}(3+11N+16N^2+10N^3+5N^4)\left(\frac{\hbar}{\Lambda}\right)^3\\
&-\frac{1}{8}(53+225N+390N^2+370N^3+165N^4+66N^5)\left(\frac{\hbar}{\Lambda}\right)^4\\
&-\frac{9}{8}(33+157N+318 N^2+350N^3+245N^4+84N^5+28N^6)\left(\frac{\hbar}{\Lambda}\right)^5\\
&+\Ocal\left(\frac{\hbar^6}{\Lambda^6}\right)\,.
\end{split}
\eq
Using the perturbative quantization condition \req{CosineQMPertQuantCond}, the function $B(\cale_p)$ can be obtained as
\beq\eqlabel{BCosineQM}
\begin{split}
B(\cale_p)=&\, \cale_p+\frac{1}{4}\left(1+4\cale_p^2\right)\frac{\hbar}{\Lambda}+\frac{\cale_p}{4}(5+12\cale_p^2)\left(\frac{\hbar}{\Lambda}\right)^2+\frac{1}{32}(17+280\cale_p^2+400\cale_p^4)\left(\frac{\hbar}{\Lambda}\right)^3\\
&+\frac{7}{64}\cale_p(103+600\cale_p^2+560\cale_p^4)\left(\frac{\hbar}{\Lambda}\right)^4\\
&+\frac{1}{256}(1619+43764\cale_p^2+129360\cale_p^4+84672\cale_p^6) \left(\frac{\hbar}{\Lambda}\right)^5+\Ocal\left(\frac{\hbar^6}{\Lambda^6}\right)\,.
\end{split}
\eq
The expansion of $B(\cale_p)$ is for $\Lambda=1$ in agreement with the expansion previously obtained in \cite{ZJJ04a,ZJJ04b}.

Solving the relation \req{EArelation} for $A(N)$ using as input \req{EpQM} leads to 
\beq\eqlabel{ACosineQM}
\begin{split}
A(N)=&\,\frac{\Lambda}{\hbar}-(2N+1)\log\Lambda+\frac{3}{2}(1+2N+2N^2)\frac{\hbar}{\Lambda}\\
&+\frac{1}{4}(11+32N+30N^2+20N^3)\left(\frac{\hbar}{\Lambda}\right)^2\\
&+\frac{5}{8}(15+52N+74N^2+44N^3+22N^4)\left(\frac{\hbar}{\Lambda}\right)^3\\
&+\frac{9}{32}(157+636N+1050N^2+980N^3+420N^4+168N^5)\left(\frac{\hbar}{\Lambda}\right)^4+ \Ocal\left(\frac{\hbar^5}{\Lambda^5}\right)\,,
\end{split}
\eq
or, in terms of $\cale_p$, 
\beq
\begin{split}
A(\cale_p)=&\,\frac{\Lambda}{\hbar}-2 B(\cale_p)\log \Lambda+\frac{1}{4}(3+12\cale_p^2)\frac{\hbar}{\Lambda}+\frac{1}{4}(23\cale_p+44\cale_p^3)\left(\frac{\hbar}{\Lambda}\right)^2\\
&+\frac{1}{64}(215+2728\cale_p^2+3184 \cale_p^4)\left(\frac{\hbar}{\Lambda}\right)^3+\frac{1}{64}(4487+20864\cale_p^3+16336\cale_p^5)\left(\frac{\hbar}{\Lambda}\right)^4\\
&+\Ocal\left(\frac{\hbar^5}{\Lambda^5}\right)\,.
\end{split}
\eq
This expansion is for $\Lambda=1$ in agreement with the one previously given in \cite{ZJJ04a,ZJJ04b}, confirming the validity of \req{EArelation}.

\paragraph{$SU(2)$: Weak coupling}

Similar as for the Sine-Gordon model discussed above, we first have to infer the relation between $r$ and the flat coordinate $a$ near weak coupling. We know that at weak coupling (using the normalization of \cite{SW94}).
$$
u(a)=\frac{a^2}{2}+\Ocal\left(\Lambda\right)\,.
$$
The expansion of $\alpha$ given in \req{alphaSmallq} and the relation \req{MAtoSU2QM} between $\alpha$ and $u$ then tells us that
$$
\boxed{
r=\frac{2a}{\hbar}
}\,.
$$
Hence, we deduce from the recursive relations \req{Adef} that 
\beq\eqlabel{uaSU2}
\begin{split}
u(a) =&\, \frac{a^2}{2}+\frac{\Lambda^4}{4a^2-\hbar^2}+\frac{(20a^2+7\hbar^2)\Lambda^8}{4(a^2-\hbar^2)(4a^2-\hbar^2)^3}+\Ocal\left(\Lambda^{12}\right)\,.
\end{split}
\eq
Using Matone's relation \req{qMatoneRelation} we can integrate \req{uaSU2} to obtain the part of the free energy, $\Fcal_\Lambda$, depending on $\Lambda$, \ie, \beq\eqlabel{FSU2weak}
\Fcal_\Lambda(a)=-\left(a^2 -\frac{\hbar^2}{24}\right)\log\Lambda-\frac{\Lambda^4}{2(4a^2+\hbar^2)}-\frac{(20a^2+7\hbar^2)\Lambda^8}{16((a^2-\hbar^2)(4a^2-\hbar^2)^3}+\Ocal\left(\Lambda^{12}\right)\,.
\eq
Under matching of normalization conventions (a rescaling of all parameters by a factor of two), the part of the free energy \req{FSU2weak} regular in $\Lambda$ is in accord with the one obtainable from the instanton counting scheme of \cite{N02}, and the $\log\Lambda$ term with \req{gammaeAsymp}. From \req{FSU2weak} we infer the first few orders in $\Lambda$ of the period $\Pi_\Lambda$ to be given by
\beq\eqlabel{PiLweakExpansion}
\Pi_\Lambda(a) = -2a\log\Lambda +\frac{4a\Lambda^4}{(4a^2-\hbar^2)^2}+\frac{3a(80a^4-16a^2\hbar^2-37\hbar^4)\Lambda^8}{8(a^2-\hbar^2)^2(4a^2-\hbar^2)^4}+\Ocal\left(\Lambda^{12}\right)\,.
\eq

\vspace{0.25cm}
\paragraph{$SU(2)$: Strong coupling}

We can infer from \cite{SW94} that the local flat coordinate $a_D$ near $u_p=-\Lambda^2$ reads
\beq\eqlabel{aDuLeadTerm}
 a_D(\t u_p)/\Lambda=\frac{\t u_p}{2\Lambda^2}+\Ocal\left(\frac{\t u_p^2}{\Lambda^4}\right)\,,
\eq
where we defined $\t u_p:= u_p+\Lambda^2$, as in \req{upDef}. With help of the relations \req{MAtoSU2QM} we deduce
$$
\boxed{
r=\frac{1}{\hbar}\left(2a_D-\frac{\hbar}{2}\right)
} \,. 
$$
Hence, we infer from $\alpha(r,q)$ calculated via \req{alphaExpansion} and \req{Gdef} that
\beq\eqlabel{uaDSU2}
\begin{split}
\t u_p(a_D)=&\,2 a_D \Lambda-\frac{1}{64}(16a_D^2+\hbar^2)-\frac{a_D(16a_D^2+3\hbar^2)}{512\Lambda}\\
&-\frac{1280a_D^4+544a_D^2\hbar^2+9\hbar^4}{131072\Lambda^2}
-\frac{8448a_D^5+6560a_D^3\hbar^2+405a_D\hbar^4}{2097152\Lambda^3}\\
&-\frac{9(14336a_D^6+17920a_D^4\hbar^2+2616a_D^2\hbar^4+27\hbar^6)}{67108864\Lambda^4}+\Ocal\left(\frac{1}{\Lambda^5}\right)\,.
\end{split}
\eq
The above $\t u_p$ is in agreement with previously obtained expansions near the dyon point (\cf, \cite{HM10a}). Note that we can as well recover \req{EpQM} from \req{uaDSU2} via combining \req{SU2QMtoCosineQM} and \req{aDuLeadTerm}.  In particular, we have the identification of flat coordinates (up to a rescaling of $\hbar$)
\beq\eqlabel{aDB}
\boxed{
a_D(\t u_p)= \frac{\hbar}{2} B(\cale_p)
}\,.
\eq
Invoking the Matone relation \req{qMatoneRelation}, we can integrate \req{uaDSU2} to obtain
\beq\eqlabel{SU2FL}
\begin{split}
\Fcal_\Lambda(a_D)=&- 8a_D\Lambda+\left( a_D^2-\frac{\hbar^2}{24}\right) \log\Lambda - \frac{a_D(16a_D^2+3\hbar^2)}{128\Lambda}\\
&-\frac{1280a_D^4+544a_D^2\hbar^2+9\hbar^4}{65536\Lambda^2}-\frac{8448a_D^4+6560a_D^2\hbar^2+405\hbar^4}{1527864\Lambda^3}\\
&-\frac{9(14336a_D^6+17920a_D^4\hbar^2+2616a_D^2\hbar^4+27\hbar^6)}{67108864\Lambda^4}+\Ocal\left(\frac{1}{\Lambda^5}\right)\,,
\end{split}
\eq
and
\beq\eqlabel{PiSU2Lambda}
\begin{split}
\Pi_\Lambda(a_D)=&\,-8\Lambda +2a_D \log\Lambda-\frac{3(16a_D^2+\hbar^2)}{128\Lambda}-\frac{80a_D^3+17a_D\hbar^2}{1024\Lambda^2}\\
&-\frac{5(2816a_D^4+1312a_D^2\hbar^2+27\hbar^4)}{524288\Lambda^3}-\frac{9(5376a_D^5+4480a_D^3\hbar^2+327a_D\hbar^4)}{4194304\Lambda^4}\\
&+\Ocal\left(\frac{1}{\Lambda^5}\right)\,.
\end{split}
\eq
Under matching of conventions, \ie, $\hbar\rightarrow 2\hbar$, the above expansion of $\Fcal_\Lambda$ is in accord with the $\Lambda$-independent part of the free energy expanded near the dyon point in moduli space inferable from \cite{KW10a}, and the $\log\Lambda$ term with \req{DeltaAsymp}. 

Making use  of relation \req{SU2QMtoCosineQM} we can also recover \req{ACosineQM} from \req{PiSU2Lambda}, \ie,
\beq\eqlabel{SU2strongAviaP}
A(N)=-\frac{2}{\hbar}\Pi_\Lambda\left(\frac{\hbar}{2}\left(N+\frac{1}{2}\right)\right)\,,
\eq
(up to a rescaling of $\hbar$).

\section{Contribution of massless vector/hyper-multiplets}
\label{appMasslessMultiplets}
\subsection{Vector-multiplet}
\label{vectorContribution}
According to \cite{NO03,NY03}, the contribution of a massless vector-multiplet to the gauge theory free energy is given by
\beq\eqlabel{gammaeeDef}
\gamma_{\ep_1,\ep_2}(x,\Lambda):=\left.\frac{d}{ds}\right|_{s=0}\frac{\Lambda^s}{\Gamma(s)}\int_0^\infty dt\, t^{s-1}\frac{e^{-tx}}{(e^{\ep_1 t}-1)(e^{\ep_2 t}-1)}\,.
\eq
We will refer to the surviving contribution in the Nekrasov-Shatashvili limit as $\gamma_{\ep_1}$, \ie,
\beq\eqlabel{gammaeDef}
\gamma_{\ep_1}(x,\Lambda):=\lim_{\ep_2\rightarrow 0}\ep_1\ep_2\,\gamma_{\ep_1,\ep_2}(x,\Lambda)\,.
\eq
It follows that 
\beq\eqlabel{gammaeInt}
\gamma_{\ep_1}(x,\Lambda)=\ep_1^2\left.\frac{d}{ds}\right|_{s=0}\frac{\Lambda^s}{\ep_1^s\Gamma(s)}\int_0^\infty dt\, t^{s-2}\frac{e^{-t (1+x/\ep_1)}}{(1-e^{-t})}\,.
\eq
The integral expression for $\gamma_{\ep_1}$ in \req{gammaeInt} can be evaluated exactly. For that, we first have to analytically continue the integral to the domain $\Re s>-1$, making use of the expansion $\frac{1}{1-e^{-x}}=\frac{1}{x}+\frac{1}{2}+\frac{x}{12}+\Ocal(x^3)$ and the integral formula
\beq\eqlabel{GammaintRep}
\int_0^\infty x^{s-1}e^{-zx}dx =z^{-s}\Gamma(s)\,,
\eq
valid for $\Re s>0$ and $\Re z>0$. This leads to
$$
\gamma_{\ep_1}(x,\Lambda)=\ep_1^2\left.\frac{d}{ds}\right|_{s=0} \left(\frac{\Lambda}{\ep_1}\right)^s\left( \eta^I(s,1+x/\ep_1)+\eta^{II}(s,x) \right)\,,
$$
with
$$
\eta^I(s,z)=\frac{z^{1-s}}{2(s-1)}+\frac{z^{-s}}{12}+\frac{z^{2-s}}{(s-1)(s-2)}\,,
$$
and
$$
\eta^{II}(s,z)=\frac{1}{\Gamma(s)}\int_0^\infty dt\, t^{s-2} e^{-t(1+x/\ep_1)}\left(\frac{1}{1-e^{-t}}-\frac{1}{t}-\frac{1}{2}-\frac{t}{12}\right)\,.
$$
Applying the derivative then leads to
$$
\left.\frac{d}{ds}\right|_{s=0} \left(\frac{\Lambda}{\ep_1}\right)^s \eta^I(s,z)= -\frac{z}{2}+\frac{3z^2}{4}-\frac{1}{2}B_2(z)\log\left(\frac{\ep_1 z}{\Lambda}\right)\,,
$$
and
$$
\left.\frac{d}{ds}\right|_{s=0} \left(\frac{\Lambda}{\ep_1}\right)^s \eta^{II}(s,z)=\frac{1}{12}-\frac{z^2}{4}+\frac{1}{2}B_2(z)\log z-\zeta'(-1,z)\,,
$$
where we made use of an identity for the derivative of the Hurwitz-Zeta function, $\zeta'(z,q):=\partial_z \zeta(z,q)$, of \cite{A03a}. 

We conclude that
\beq\eqlabel{gammaeResult}
\boxed{
\gamma_{\ep_1}(x,\Lambda)=\frac{\ep_1^2}{2}B_2(1+x/\ep_1)\left(1+\log\left(\frac{\Lambda}{\ep_1}\right)\right)-\ep_1^2\,\zeta'(-1,1+x/\ep_1)
}\,.
\eq
Note that $\gamma_{\ep_1}$ can be easily asymptotically expanded for large $x$ using the relation (see for instance \cite{EM01})
$$
\zeta'(-1,1+z)=\zeta'(-1,z)+z \log z\,,
$$
and making use of the known asymptotic expansion of $\zeta'(-1,z)$ (see \cite{E86,R89}),
$$
\zeta'(-1,z)\sim\frac{1}{12}-\frac{1}{4}z^2+\frac{1}{2}B_2(z)\log z-\sum_{k=1}^\infty \frac{B_{2k+2}}{2k(2k+1)(2k+2)}z^{-2k}\,,
$$
with $B_k$ denoting the $k$th Bernoulli number. Hence,
\beq\eqlabel{gammaeAsymp}
\begin{split}
\gamma_{\ep_1}(x;\Lambda)\sim&\, \frac{x^2}{4}\left(3+2\log\left(\frac{\Lambda}{x}\right)\right)+\frac{x}{2}\left(1+\log\left(\frac{\Lambda}{x}\right)\right)\ep_1+\frac{1}{12}\log\left(\frac{\Lambda}{x}\right)\ep_1^2\\
&+\ep_1^2\sum_{k=1}^\infty \frac{B_{2k+2}}{2k(2k+1)(2k+2)}\left(\frac{\ep_1}{x}\right)^{2k}\,.
\end{split}
\eq
The resulting expansion is in perfect agreement with the limit \req{gammaeDef} of the asymptotic expansion obtained for $\gamma_{\ep_1,\ep_2}$ in \cite{NY03}. 

The derivative of $\gamma_{\ep_1}$ can be obtained via making use of the identities,
$$
\frac{\partial}{\partial a} \zeta(s, a)=-s\,\zeta(s+1,a)\,,
$$
and
\beq
\begin{split}
\zeta(0,a)&=\frac{1}{2}-a\,,\\
\zeta'(0,a)&=\log\Gamma(a)-\frac{1}{2}\log 2\pi\,.
\end{split}
\eq
We infer
\beq\eqlabel{Dgammae}
\boxed{
\frac{\partial\gamma_{\ep_1}(x,\Lambda)}{\partial x}=\ep_1\left(\frac{1}{2}+\frac{x}{\ep_1}\right)\log\left(\frac{\Lambda}{\ep_1}\right)-\ep_1\log\Gamma\left(1+x/\ep_1\right)+\frac{\ep_1}{2}\log 2\pi
}\,.
\eq
An asymptotic expansion of the above formula can be obtained by making use of the classical asymptotic expansion of $\log\Gamma$, (see \cite{NIST10}), 
\beq\eqlabel{LogGasymp}
\log\Gamma(h+z)\sim(z+h-1/2)\log z-z+\frac{1}{2}\log 2\pi+\sum_{k=2}^\infty \frac{(-1)^kB_{k}(h)}{k(k-1)}\, z^{1-k}\,,
\eq
for $z\rightarrow\infty$ with $|\arg z|<\pi$, $h\in [0,1]$ and where $B_k(h)$ denotes the $k$th Bernoulli polynomial. In particular, $B_k(1)=B_k$ is the $k$th Bernoulli number.

\subsection{Hyper-multiplet}
The contribution of a hypermultiplet, denoted as $\delta_{\ep_1,\ep_2}(x,\Lambda)$, is given by a simple shift of \req{gammaeeDef} (see \cite{GNY10,KW10b})
\beq\eqlabel{deltaeeDef}
\delta_{\ep_1,\ep_2}(x,\Lambda)=\gamma_{\ep_1,\ep_2}(x-(\ep_1+\ep_2)/2,\Lambda)\,.
\eq
Correspondingly, under the definition 
$$
\delta_{\ep_1}(x,\Lambda):=\lim_{\ep_2\rightarrow 0}\ep_1\ep_2\,\delta_{\ep_1,\ep_2}(x,\Lambda)\,,
$$
we immediately deduce from \req{gammaeResult} and \req{Dgammae} that 
\beq\eqlabel{Deltaep}
\boxed{
\delta_{\ep_1}(x,\Lambda)=\frac{\ep_1^2}{2}B_2\left(\frac{1}{2}+x/\ep_1\right)\left(1+\log\left(\frac{\Lambda}{\ep_1}\right)\right)-\ep_1^2\,\zeta'\left(-1,\frac{1}{2}+x/\ep_1\right)
}\,,
\eq
and
\beq\eqlabel{Ddeltae}
\boxed{
\frac{\partial \delta_{\ep_1}(x,\Lambda)}{\partial x}=x\log\left(\frac{\Lambda}{\ep_1}\right)-\ep_1\log\Gamma\left(\frac{1}{2}+x/\ep_1\right)+\frac{\ep_1}{2}\log 2\pi
}\,.
\eq
The asymptotic expansion of \req{Ddeltae} can be obtained with help of \req{LogGasymp}, where now $h=1/2$, \ie,
\beq\eqlabel{DDeltaAsympt}
\frac{\partial \delta_{\ep_1}(x,\Lambda)}{\partial x}\sim x\log\left(\frac{\Lambda}{x}\right)+x-\ep_1\sum_{k=2}^\infty\frac{(-1)^kB_k(1/2)}{k(k-1)} \left(\frac{\ep_1}{x}\right)^{k-1}\,.
\eq
However, we are not aware that the general asymptotic expansion of $\zeta'\left(-1,h+z\right)$ with $h\in [0,1]$ has been derived in the mathematics literature. Hence, in order to infer the asymptotic expansion of $\delta_{\ep_1}$ we proceed similar as in \cite{NY03}, \ie, we explicitly integrate \req{gammaeInt} after expanding the (shifted) integrand. For that, recall that the generating function of Bernoulli polynomials reads
\beq\eqlabel{GenFBernoulliP}
\frac{t e^{ht}}{e^t-1}=\sum_{n=0}^\infty B_n(h) \frac{t^n}{n!}\,.
\eq
We infer from \req{gammaeInt} with \req{GenFBernoulliP} and \req{GammaintRep} that
\beq\eqlabel{DeltaAsymp}
\begin{split}
\delta_{\ep_1}(x,\Lambda)\sim&\,\ep_1^2 \left.\frac{d}{ds}\right|_{s=0} \left(\frac{\Lambda}{x}\right)^s \sum_{n=0}^\infty B_n(1/2) \frac{\Gamma(s+n-2)}{\Gamma(1+n)\Gamma(s)}\, \left(\frac{\ep_1}{x}\right)^{n-2}\\
&= \left(\frac{x^2}{2} -\frac{1}{24} \ep_1^2 \right)\log\left(\frac{\Lambda}{x}\right)+\frac{3}{4}x^2 +\ep_1^2\sum_{n=4}^\infty\frac{ B_n(1/2) }{n(n-1)(n-2)} \left(\frac{\ep_1}{x}\right)^{n-2}\,.
\end{split}
\eq
Taking the derivative of the above asymptotic expansion of $\delta_{\ep_1}$ reproduces \req{DDeltaAsympt}.

\end{document}